\documentclass[fleqn,usenatbib]{mnras}

\usepackage[T1]{fontenc}
\DeclareRobustCommand{\VAN}[3]{#2}
\let\VANthebibliography\thebibliography
\def\thebibliography{\DeclareRobustCommand{\VAN}[3]{##3}\VANthebibliography}

\usepackage{graphicx}
\usepackage{amsmath}
\usepackage{amssymb}
\usepackage{xcolor}

\usepackage[normalem]{ulem}

\title[Magnetized galactic discs]{Rotation measure and synchrotron emission signatures in simulations of magnetized galactic discs}

\author[Y. Rappaz et al.]{
Yoan Rappaz$^{1},$\thanks{E-mail: yoan.rappaz@epfl.ch}
Jennifer Schober$^{1}$,
Philipp Girichidis$^{2}$
\\
$^{1}$Laboratoire d'astrophysique, \'Ecole Polytechnique Fédérale de Lausanne (EPFL), Observatoire de Sauverny
CH–1290 Versoix, Switzerland\\
$^{2}$Universit\"{a}t Heidelberg, Zentrum f\"{u}r Astronomie, Institut f\"{u}r Theoretische Astrophysik, Albert-Ueberle-Str. 2, D-69120 Heidelberg, Germany}

\date{Accepted XXX. Received YYY; in original form ZZZ}

\pubyear{2021}

\begin{document}
\label{firstpage}
\pagerange{\pageref{firstpage}--\pageref{lastpage}}
\maketitle

\begin{abstract}
We analyse observational signatures of magnetic fields for simulations of a Milky-Way 
like disc with supernova-driven interstellar turbulence and self-consistent chemical 
processes. 
In particular, we post-process two simulations data sets of the SILCC Project 
for two initial amplitudes 
of the magnetic field,  
$B_0 = $ 3  and 6 $\mu$G, to study the evolution of Faraday rotation measures (RM) and synchrotron luminosity.
For calculating the RM, three different models of the electron density $n_e$ are considered. 
A constant electron density, and two estimations based on 
the density of ionized species and the fraction of the total gas, respectively.
Our results show that the RM profiles are extremely sensitive to the $n_e$ models,
which assesses the importance of accurate electron distribution observations/estimations 
for the magnetic fields to be probed using Faraday RMs.
As a second observable of the magnetic field, we estimate the synchrotron luminosity in the simulations
using a semi-analytical cosmic ray model.
We find that the synchrotron 
luminosity decreases over time, which is connected to the decay of magnetic energy in the simulations.
The ratios between the magnetic, the cosmic ray, and the
thermal energy density indicate that the assumption of equipartition does not hold for most regions of the ISM. 
In particular, for the ratio of the cosmic ray to the
magnetic energy the assumption of equipatition 
could lead to a wrong interpretation of the observed synchrotron emission.
\end{abstract}



\section{Introduction}
\label{sec:introduction}
Magnetic fields are ubiquitous in the universe. 
They are observed at a wide range of spatial scales and field strengths;
the average magnetic field strength on the surface of Earth is about $0.5$~G \citep{Earth_magfield_1, Earth_magfield_2} and on the Sun about $1$ G \citep{Sun_mean_magfield_Scherrer1977}, 
while it can reach several thousand Gauss in sun spots
(e.g.~\citealt{sunspots_magfields1,sunspots_magfields}). 
Entire (spiral) galaxies are magnetized with ordered magnetic fields of typically $10^{-6}$ to $10^{-5}$ G (e.g.\ \citealt{beck_2001_magfields, Beck_2011_magnfield_milkywaylike, beck_magfields_spirals,KrauseEtAl2018}).
Galaxy clusters and the intracluster medium also show evidence of magnetic fields of the microgauss order
\citep{annual_rev_magfields_obs, StuardiEtAl2021}.
The study of magnetic fields is a central part of astrophysical research. One reason is that their influence and involvement in many processes is now well established.
For example, the magnetic fields of spiral galaxies could influence the star formation rate \citep{magfields_SFR_Krumholz2019}. They also 
play a role in cosmic ray (CR)
propagation \citep{Zweibel2013, Zweibel2017, magfields_cosmicrays}, and are a main driver of stellar activity
\citep{magfields_stellar_activity_1,magfields_stellar_activity_2}.

Magnetic field amplification in many astrophysical
systems is often explained by dynamo activity \citep{dynamo_big_reference}. 
In spiral galaxies, for instance, the $\alpha$-$\Omega$ dynamo can convert kinetic energy from to differential rotation and turbulence into magnetic energy 
\citep{beck_magfields_spirals}. 
However, magnetic fields aligned with the galaxy's
differential rotation axis have a characteristic strength of the order of one micrograuss \citep[e.g.][]{NaabOstriker2017} which cannot be explained by the $\alpha$-$\Omega$ dynamo alone since the
amplification time characteristic of this process is much longer than the age of the system itself \citep{ruzmaikin2013magnetic}. 
Therefore, additional dynamo processes should be involved to reach the observed magnetic field strengths in galaxies, such as the small-scale turbulent dynamo \citep{brandenburg_dynamo, brandenburg_current_status_dynamo, SchoberEtAl2013}.
This phenomenon amplifies magnetic fields on scales smaller than the characteristic
injection scale of turbulence that is primarily driven by 
supernova (SN) explosions 
in the interstellar medium. An exhaustive literature on the small-scale dynamo as the main mechanism for the amplification of magnetic fields in the interstellar medium is available (see for example \citealt{ssd_galaxies_example_2}, \citealt{ssd_galaxies_example_1}).

To distinguish different scenarios of magnetic field generation, reliable observational 
tracers of astrophysical magnetic fields are needed. 
Many observational techniques allow to obtain various information on the
structure and strength of the fields. 
The Faraday rotation measure (RM), for example, allows 
to determine the magnetic field's projection along the line of sight, by analyzing the rotation of the polarization plane of electromagnetic waves coming from radio sources located in the background or within the observed environment
\citep{faraday_screens_power_spectrum, faraday_spread_function, faraday_rotation_SSD}. 
The measurement of polarized synchrotron emission makes it possible to study the component normal to the line of sight, as well as to estimate the total intensity of the field \citep{Orlando_synchrotron_magfields}. However, since the observed quantities are usually field projections, the reconstruction of 
the exact spatial structure of the magnetic field remains a significant challenge (e.g. \citealt{magfields_challenge, SeifriedEtAl2020, Girichidis2021}). Indeed, the computation of the RM requires to know the distribution of thermal electrons, and the estimation of the three-dimensional structure of the field from radio observations requires to measure the Stokes parameters and to involve Faraday tomography \citep[e.g.][]{rotation_measure_synthesis}.
Synchrotron emission is emitted by relativistic charged 
particles that travel in the interstellar magnetic field is the second observable. 
However, extracting the magnetic field from synchrotron observation
requires an assumption on the power spectrum of the cosmic rays 
\citep[e.g.][]{blumenthal_radiation}. 

Despite a constant improvement of the theoretical tools describing the dynamics of 
astrophysical plasmas, difficulties are encountered in the attempt to match observations 
to theoretical models. 
The need for detailed numerical modelling of the 
interstellar medium 
(ISM) then becomes a necessity.
Some of the most detailed simulations of the
magnetized turbulent Milky Way ISM are presented in the SILCC project\footnote{Official webpage: \url{https://hera.ph1.uni-koeln.de/~silcc/}} 
\citep{WalchEtAl2015, GirichidisEtAl2016b}.
The central aim of the SILCC simulations is the 
analysis of the formation and evolution of molecular clouds, as well as their dynamical
characteristics. 
Specifically, they carry out an in-depth study on the influence of stellar feedback, like supernovae, on the evolution of the ISM. 
The magnetohydrodynamical simulations cover the ISM in different phases (ionized, atomic and molecular) from $10^6$\,K down to $10$\,K with densities ranging from $10^{-4}-10^4\,\mathrm{cm^{-3}}$. \citet{SILCC_project_molecularclouds} (data release 6, DR6) focus on the magnetic field, which allows a detailed comparison of the field properties in different phases of the ISM \citep[see also][for comparable models in periodic boxes of the local SN-driven ISM]{PardiEtAl2017}.

The objective of this paper is the implementation and analysis of two observables of the magnetic field to the SILCC simulation set, in order to probe the sensitivity of typical assumption made in the reconstruction of magnetic fields from radio observations. In a first step, we will study the dynamics of the Faraday rotation over the simulation time for different models of the electron density.
In a second step, we will use the model for the 
cosmic-ray electrons developed by \cite{schober_galactic_synchrotron_emission}, 
to post-process the simulation data, 
in order to study the time evolution of the synchrotron luminosity.
Additionally, we will calculate the energy density of cosmic rays and magnetic fields as well as the thermal energy
in order to test the equipartition between these components.
The structure of this paper is organized as follows. In Sec.~\ref{sec:simulations_data}, we briefly review the simulations from the SILCC project.
In Sec.~\ref{sec:faraday_rotation}, we compute Faraday rotation measures for the simulations from the SILCC project (DR6), in particular by implementing different distribution models for the 
free electrons. 
In Sec.~\ref{sec:SYNCH_emission}, we present the synchrotron emission model applied to the DR6 simulations, 
and the evolution of the energy densities. 
Conclusions and perspectives are discussed in Sec.~\ref{sec:conclusions}.

\section{Simulation data}
\label{sec:simulations_data}

\subsection{The DR6 simulation set}
\label{subsec:SIMDATA_DR6_simulations}

We post-process simulations from the SILCC project (SImulating the Life Cycle of molecular 
Clouds; \citealt{WalchEtAl2015, GirichidisEtAl2016b}). 
The physics modules and the simulation setup are described in detail 
in \citet{WalchEtAl2015}. The effects of magnetic fields have been investigated in \citet{SILCC_project_molecularclouds} and were published as data release 6 (DR6). Below we summarize the main simulation features.

The simulation setup covers a stratified box with a size of $0.5\times0.5\times0.5\,\mathrm{kpc}^3$. The setup uses periodic boundary conditions along $x$ and $y$ and diode boundary conditions along the stratification axis $z$, which means the gas can leave but not enter the simulation box. The equations of ideal MHD are solved using the HLLR5 solver \citep{Bouchut2007, Bouchut2010, Waagan2009, Waagan2011} in the adaptive mesh refinement (AMR) code \textsc{Flash}\footnote{Version 4, \url{http://flash.uchicago.edu/site/}} (\citealt{FLASH00,DubeyEtAl2008}). Radiative cooling and heating of the gas are coupled to the evolution of the non-equilibrium abundances using a chemical network. The hydrogen and carbon chemistry including the abundances of ionized (H$^+$), atomic (H) and molecular (H$_2$) hydrogen as well as carbon monoxide (CO) and singly ionized carbon
follows \citet{GloverMacLow2007a}, \citet{MicicEtAl2012}, and 
\citet{NelsonLanger1997}. Molecular cooling follows the description in \citet{GloverEtAl2010} and \citet{GloverClark2012b}. For temperatures above $10^4\,\mathrm{K}$, equilibrium abundances are assumed and the cooling based on \citet{GnatFerland2012} is employed. Heating of the gas includes spatially clustered supernovae (SNe), which is described in more detail below, cosmic ray \citep{GoldsmithLanger1978} and X-ray \citep{WolfireEtAl1995} heating as well as photoelectric heating \citep{BakesTielens1994, Bergin2004, WolfireEtAl2003}. The constant interstellar radiation field with a strength of $G_0=1.7$ \citep{Habing1968, Draine1978} is locally attenuated in dense shielded regions. The local column depth is computed using the TreeCol algorithm \citep{ClarkGloverKlessen2012}, which has been implemented and optimized for \textsc{Flash} as described in \citet{WuenschEtAl2018}. The dust-to-gas mass ratio is set to 0.01 with dust opacities based on \citet{MathisMezgerPanagia1983} and \citet{OssenkopfHenning1994}.

The initial gas distribution follows a Gaussian profile 
with a scale height of $30\,\mathrm{pc}$ and a central density of $\rho_0=9\times10^{-24}\,\mathrm{g\,cm}^{-3}$, which yields a total gas column density of $\Sigma=10\,\mathrm{M}_\odot\mathrm{pc}^{-2}$. The temperature is adjusted such that the gas is initially in pressure equilibrium. This corresponds to a central temperature $T$ of $4600\,\mathrm{K}$ and a value of $T=4\times10^8\,\mathrm{K}$ at large altitudes, where a lower boundary to the density ($\rho_\mathrm{min}=10^{-28}\,\mathrm{g\,cm}^{-3}$) is applied.

Besides self-gravity an external potential is included that accounts for the stellar component of the disc. An isothermal sheet is used \citep{Spitzer1942} with a stellar gas surface density of $30\,\mathrm{M}_\odot\mathrm{pc}^{-2}$ and a vertical scale height of $100\,\mathrm{pc}$. The gravitational forces are computed using the tree-based method described in \citet{WuenschEtAl2018}.

The initial magnetic field is oriented along the $x$ direction. The central field strength in the midplane at $z=0$ is set to $B_0=3\,\mu\mathrm{G}$ in one simulation and to $B_0=6\,\mu\mathrm{G}$ in a second run and scales vertically with the density,
$B_x(z)=B_0\,[\rho(z)/\rho(z=0)]^{1/2}$. 
An initial small scale tangled perturbation of the field is not included. The mass-to-flux ratio (measured along the $x$ direction) of the entire box is approximately $(M/\Phi)/\mu_\mathrm{crit}=6.72$ in units of the critical value $\mu_\mathrm{crit}=(2\pi G^{1/2})^{-1}$, i.e., the disc as a whole is not supported by the magnetic pressure.
The initial conditions of a field parallel to the midplane is motivated by observations of the Milky Way.

Stellar feedback is included as a temporally constant SN rate. The Kennicutt-Schmidt relation \citep{KennicuttSchmidt1998} is used to find a star formation rate based on the total gas column density of the simulation box. The star formation rate is then converted to a SN rate assuming the initial stellar mass function from \citet{Chabrier2003}. This results in a rate of approximately $15\,\mathrm{Myr}^{-1}$ for the box. A distinction is made between a type~Ia component (20\% of the SNe) with a uniform distribution 
in $x$ and $y$ directions and a Gaussian distribution in $z$ direction
with a scale height of $300\,\mathrm{pc}$ and a type~II component (the remaining 80\% of the SNe) with a vertical scale 
height of $50\,\mathrm{pc}$. 
The latter are further split into a runaway component (40\%) of individual explosions and a clustered component (60\%), which are grouped in clusters with sizes ranging from 7 to 40 SNe per cluster. We note that this simplified treatment of stellar feedback is lacking self-consistent star formation and additional direct feedback from localised star clusters like stellar winds and radiation \citep[see, e.g.][]{GattoEtAl2017, RathjenEtAl2021}. However, the inclusion of 
Lagrangian sink particles 
as a representation of star clusters to accurately follow the star formation and feedback includes an ad-hoc treatment of the magnetic field for the accretion of gas from the grid to the particles. We therefore restrict our analysis to these simplified models.

The grid is initialized at a resolution of $128^3$ cells. The adaptive mesh refinement further increases the resolution by a factor of four, i.e. an effective resolution of $512^3$ cells with a cell size of $\Delta x = 0.98\,\mathrm{pc}$. 
For the post-processing, we extract the data as a uniform grid at the highest resolution, namely $512^3$ grid cells.
We use $13$ data snapshots from $0$ to $60$ Myr with a constant time step of $5$ Myr. 
A discussion about the effect of resolution on 
the final results of our analysis is presented in
Appendix~\ref{appendix:varying_resolution}.

\subsection{Magnetic field evolution}
\label{subsubsec:FRM_magfield_evolution}

\begin{figure}
\begin{center}
	\includegraphics[width=\columnwidth]{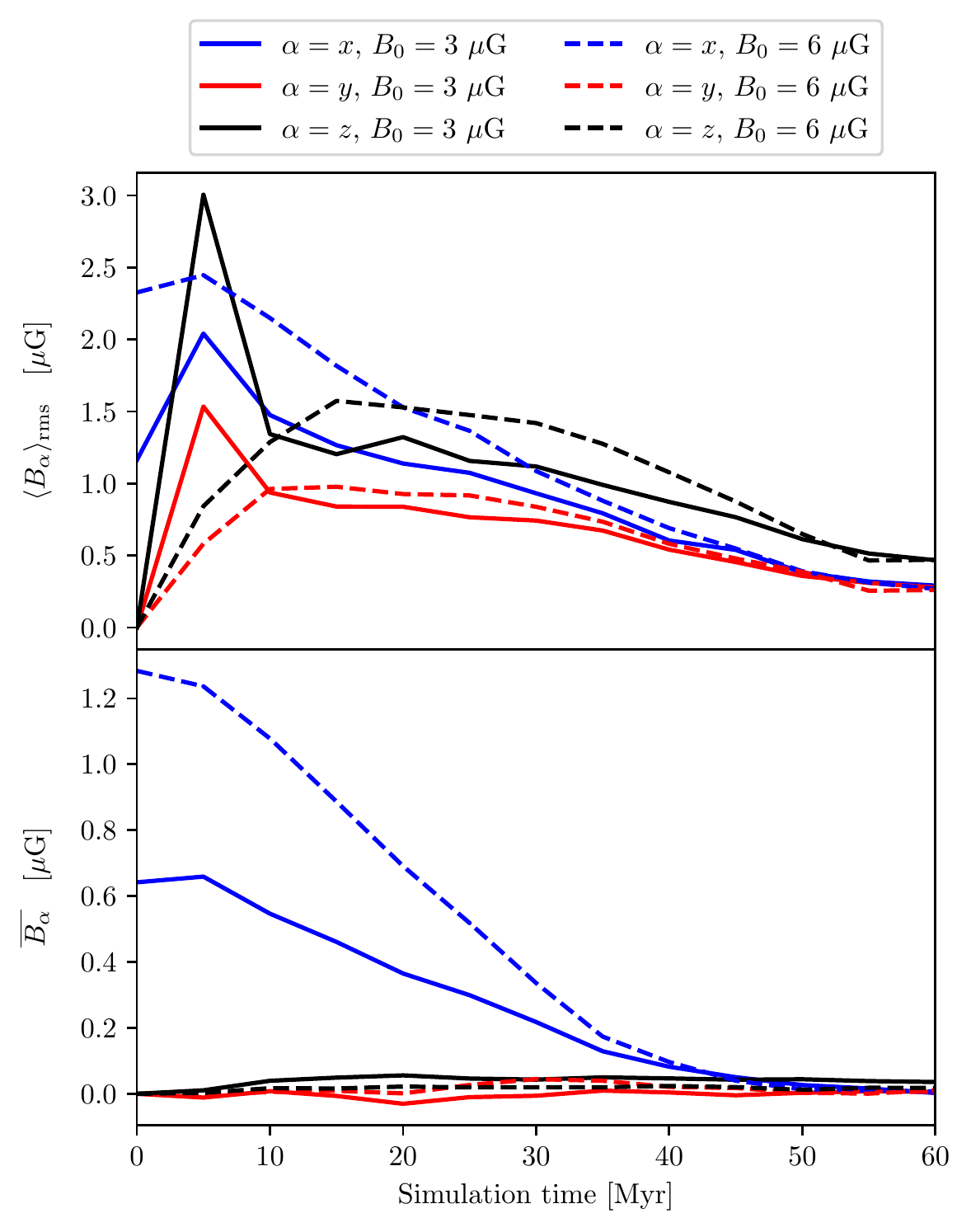}
    \caption{
    Time evolution of the volume-weighted rms value (top) and plane average (bottom)
    of different magnetic field components in the simulations with $B_0=3~\mu\mathrm{G}$ (solid lines) and $B_0=6~\mu\mathrm{G}$ (dashed lines).
}
\label{fig:RM_magfields_evolution}
\end{center}
\end{figure}

Figure~\ref{fig:RM_magfields_evolution} 
shows the time evolution of the 
different components of the magnetic field 
for the two simulations with initial volume-weighted strengths of
$B_0 = 3$ and 6 $\mu$G, respectively. 
Initially, the magnetic field increases mainly due to adiabatic compression 
of the field caused by supernovae explosions and gravitational contraction. 
The $y$ and $z$ components of the magnetic field
are initially zero. 
In both runs, $\langle B_x \rangle_\mathrm{rms}$ 
increases initially, but then decays after $5$ Myr. 
The components $\langle B_y \rangle_\mathrm{rms}$ and 
$\langle B_z \rangle_\mathrm{rms}$ show a similar evolution 
for the run with $B_0 = 3$ $\mu$G, 
but in the run with $B_0 = 6$ $\mu$G they are 
amplified in the first $15$ Myr, before decaying.
On the other hand, we observe that the arithmetic mean $\overline{B}_x$, decreases 
progressively for the two different values of $B_0$ and that $\overline{B}_y$ and $\overline{B}_z$ are close to zero over the entire simulation time.

These different behaviors can be explained by the following arguments. 
\citet{SILCC_project_molecularclouds} have shown that the magnetic field 
amplitude scales with the total gas density $\rho$ in the following way: 
for low density and weak magnetic fields
$B \propto \rho^{2/3}$, while for zones with a density close to the mean density of the 
ISM ($\rho \sim 10^{-24}$ g cm$^{-3}$),
$B \propto \rho^{1/4}$. 
Since supernova-driven turbulence creates dense zones via converging flows, an amplification of the magnetic energy then becomes 
consistent with the evolution of the rms average of the different components of the field. 
The fact that $\overline{B}_y$ and $\overline{B}_z$ oscillate close to 0 can be explained by the isotropy of the field, caused by randomly placed supernovae.
Finally, the decay of magnetic energy, 
mainly due to the combination of supernovae explosions that tend to disperse 
gas across the domain, is also in line with our 
interpretation. 
Figure~\ref{fig:mass_energies_numbparticles_evolution} 
in the appendix shows the evolution of the individual energies for completeness as well as the total mass in the simulation box. 
For a detailed discussion of the energies and the outflows we refer the
reader to \citet{SILCC_project_molecularclouds}.



\section{Faraday rotation measures}
\label{sec:faraday_rotation}

\subsection{Methodology}
\label{subsec:FRM_methodology}

\subsubsection{Basic equations}
\label{subsec:FRM_basic_equations}

A linearly polarized electromagnetic wave travelling through 
a medium with a characteristic size $L$ and a magnetic field parallel to 
the propagation of the wave $B_{\parallel}$ undergoes rotation of its polarization plane with a characteristic rotation angle 
\begin{equation}\label{equ:faraday_rotation_angle}
\Phi(\lambda^2) = \Phi_0 + \text{RM}~\lambda^2.
\end{equation}
Here, $\lambda$ is the wavelength, $\Phi_0$ is the initial angle of the polarization plane, and
\begin{equation}
\label{eq:FRM_RM_definition}
\text{RM} \equiv K\int_{0}^{L}n_e(l)B_{\parallel}(l) ~\mathrm{d}l 
\end{equation}
is called the \textit{rotation measure} (or \textit{Faraday depth}) with 
$K = e^3/(2\pi m_e^2c^4)\simeq 0.81$ rad m$^{-2}$ cm$^{3}$ $\mu$G$^{-1}$ pc$^{-1}$, 
where $e, m_e$, and $c$ are the charge and mass of the electron, and the speed of light, respectively.

In our analysis, we consider an ideal situation in the sense that each pixel of a RM map 
has a radio source in the background, and we take into account neither any broadening process, 
nor the effects of multiple sources on the Faraday spectrum 
(for more details, see for example \citealt{rotation_measure_synthesis}).

\subsubsection{Data treatment}
\label{subsubsec:FRM_data_treatment}

We use the YT software package \citep{TurkEtAl2011} to analyse the simulations.
To calculate RM, we discretize formula \eqref{eq:FRM_RM_definition} to give 
\begin{equation} 
\label{equ:FRM_rm_approx}
\text{RM} \simeq K \sum_{j = 1}^{N}B_{\parallel}(x_j)n_e(x_j)\Delta x,
\end{equation}
where $B_{\parallel}$ is the parallel component of the magnetic field to the axis along which the RM is estimated and $\Delta x$ is the size of a grid cell.
For a cubic numerical domain of dimension $N^3$, there are $N^2$ possible lines of sight (LOS) that can be calculated on the plane perpendicular to each Cartesian axis, namely $x$, $y$, and $z$. For extracting the different data, we map the original grid with adaptive mesh refinement onto a uniform grid with a resolution of $N=512^3$.
Finally, the RM is calculated according to formula \eqref{equ:FRM_rm_approx}.

Equation~\eqref{eq:FRM_RM_definition} implies that a completely random 
configuration of the magnetic field will lead to 
a (plane) average of
\begin{equation}
\label{equ:FRM_plane_formula}
    \overline{\mathrm{RM}}\equiv \frac{1}{N^2}\sum_{i = 1}^{N^2}\mathrm{RM}_i \simeq 0,
\end{equation}
where the sum is performed over all lines of sight of a given plane. 
Therefore, we will also calculate the rms value of the RM: 
\begin{equation}\label{equ:FRM_rms_formula}
   \left<\text{RM} \right>_{\text{rms}} \equiv \sqrt{\frac{1}{N^2} \sum_{i=1}^{N^2}\mathrm{RM}_{i}^2}.
\end{equation}
Obviously, the rms-calculation process could in itself present a number of drawbacks;
if zones with strong magnetic fields (or high electron density) form, they will significantly increase the RM value (a concrete example will be
presented in Sec.~\ref{subsec:FRM_results}). 
In particular, intuitive information about the structure of the field (or at least the inversion structure of the RM values) is lost, because the square of each $\text{RM}_i$ is considered (see equation \ref{equ:FRM_rms_formula}). Therefore, comparing those two averages can give us indications about structural properties of the magnetic field. Furthermore, we also consider the following estimator:
\begin{equation}\label{eq:Delta_estimator}
\Delta(\text{RM}) \equiv \max(|\text{RM}|) - \overline{\text{RM}},
\end{equation}
which measures the deviation between the highest value of RM and its plane average. 
We will use this quantity as an indicator of the (possible) 
existence of high-RM zones that could lead to an increase in 
$\left<\text{RM} \right>_{\text{rms}}$.
The definitions of the plane average as 
given in Eq.~(\ref{equ:FRM_plane_formula}), the rms as given in
Eq.~(\ref{equ:FRM_rms_formula}), and the estimator 
given in Eq.~(\ref{eq:Delta_estimator}) will be applied for other 
quantities in simulations as well. 

Another useful quantity that we analyse is called the \textit{dispersion measure} (DM) and is defined as:
\begin{equation}
    \text{DM} \equiv \int_0^L n_e(l) ~\mathrm{d}l.
\end{equation}
The DM basically gives information about the projected electron density, and could also tell us if high-value RM and DM patches are observed in the same region, which could 
indicate
that the electron density might have a greater influence on the RM than magnetic field variations. 
Averages and the $\Delta$ estimator of DM are defined in the same way as for RM.

Finally, to study typical length scales of various quantities, 
such as the RM maps and the 
free electron density,
we calculate power spectra.
For a given physical quantity $\varphi$, we define its correlation function as:
\begin{equation}
C_{\varphi}(\boldsymbol{r}) \equiv \left< \varphi(\boldsymbol{r})\varphi(\boldsymbol{x}+\boldsymbol{r})\right>,
\end{equation}
where $\boldsymbol{r}$ is the position vector between two points, and where the average is performed over all possible points $\boldsymbol{x}$. Considering an isotropic system, the latter quantity only depends on $r \equiv |\boldsymbol{r}|$. Taking the Fourier transform of the last expression, we can write:
\begin{equation}\label{power_spectrum_formula}
C_{\varphi}(r) \propto \sum_{\boldsymbol{k}} \left< |\hat{C}_{\varphi}(k)|^2\right>e^{i \boldsymbol{k}\cdot \boldsymbol{x}} \equiv \sum_{\boldsymbol{k}}P_{\varphi}(\boldsymbol{k})e^{i\boldsymbol{k}\cdot \boldsymbol{x}},
\end{equation}
where 
\begin{equation}
    \hat{C}_{\varphi}(k) \equiv \int_V d^3\boldsymbol{x}C_{\varphi}(\boldsymbol{r})e^{-i\boldsymbol{k}\cdot \boldsymbol{x}} 
\end{equation}
and where $P_{\varphi}(\boldsymbol{k})$ is the power spectrum of $\varphi$. 
The power spectrum is then averaged over all $\boldsymbol{k}$-shells with a constant
wavenumber amplitude, in order to obtain a one-dimensional 
profile\footnote{For the numerical calculation we use the python 
package \texttt{scipy.fft} \citep{Cooley:1965zz}.}. 
Note that formula \eqref{power_spectrum_formula} can be applied either to three-dimensional data, as well as weight-projected two-dimensional maps.

\subsubsection{Thermal electron density models}
\label{subsubsec:FRM_ne_models}
In this section we present three different electron density models that are implemented in the calculation of the rotation measure. The objective of considering multiple density models is to test the response of the Faraday signal, and then establish if offset RM values are observed between, for example, a basic model of constant electron density and a more complex model, accounting for the gas dynamics of the ISM. 
Assuming a constant electron density for performing an RM 
analysis 
is not uncommon in the literature, especially in numerical 
simulations. 
We can cite, as an example, the (nearly) incompressible simulations conducted by \citealt{faraday_rotation_SSD} in order to study the Faraday signal in the framework of the small-scale dynamo, assuming a constant electron density because of the small variations of the latter. So in order to test the effect of more complex distribution of electron density on the RM, we will consider the following models:

\begin{itemize}
    \item {\textit{$n_e$ model 1}. First, we consider the most trivial situation where the thermal electron distribution is constant, and where its value is set to $n_e = 10^{-3}$ cm$^{-3}$. Given formula \eqref{equ:FRM_rm_approx}, only the magnetic fields should directly influence the rotation measure.
}
    \item {\textit{$n_e$ model 2}. Here, the thermal electron density is computed via the mass fraction of the ionized 
species implemented in the simulations (C$^+$ and H$^+$).  
If $f_{\mathrm{C}+}$ is the carbon mass fraction with respect to the total gas density, the 
contribution to electron density from ionized carbon will be given as 
$n_{\mathrm{C}+} = f_{\mathrm{C}+}/m_\mathrm{C}$, where $m_\mathrm{C}$
is the mass of carbon in atomic mass units. 
The total electron density will be
given by $n_e = n_{C+} + n_{H+}$ (where $n_{H+}$ is calculated with the same procedure).
}
    \item {\textit{$n_e$ model 3}. 
    Lastly, we make the hypothesis that the electron density can be written in
    the form $n_e \propto n$ or equivalently $n_e = fn$, where $n$ is the total particle density expressed in cm$^{-3}$, and $f$ is a proportionality factor. In particular, $n$ is obtained by dividing the mass density $\rho$ times the mass fraction $\alpha_i$ of each species $i$ by their
    corresponding mass $m_i$, i.e.:
    \begin{equation}
    n \equiv  \sum_i\frac{\alpha_i \rho}{m_i}.
    \end{equation}
    In this work, we will use $f = 10^{-3}$ which results in values 
    of $n_e$ that are, on average, of the same order of magnitude as the ones obtained 
    from $n_e$ models 1 and 2. 
}
\end{itemize}

\begin{figure}
\begin{center}
	\includegraphics[width=\columnwidth]{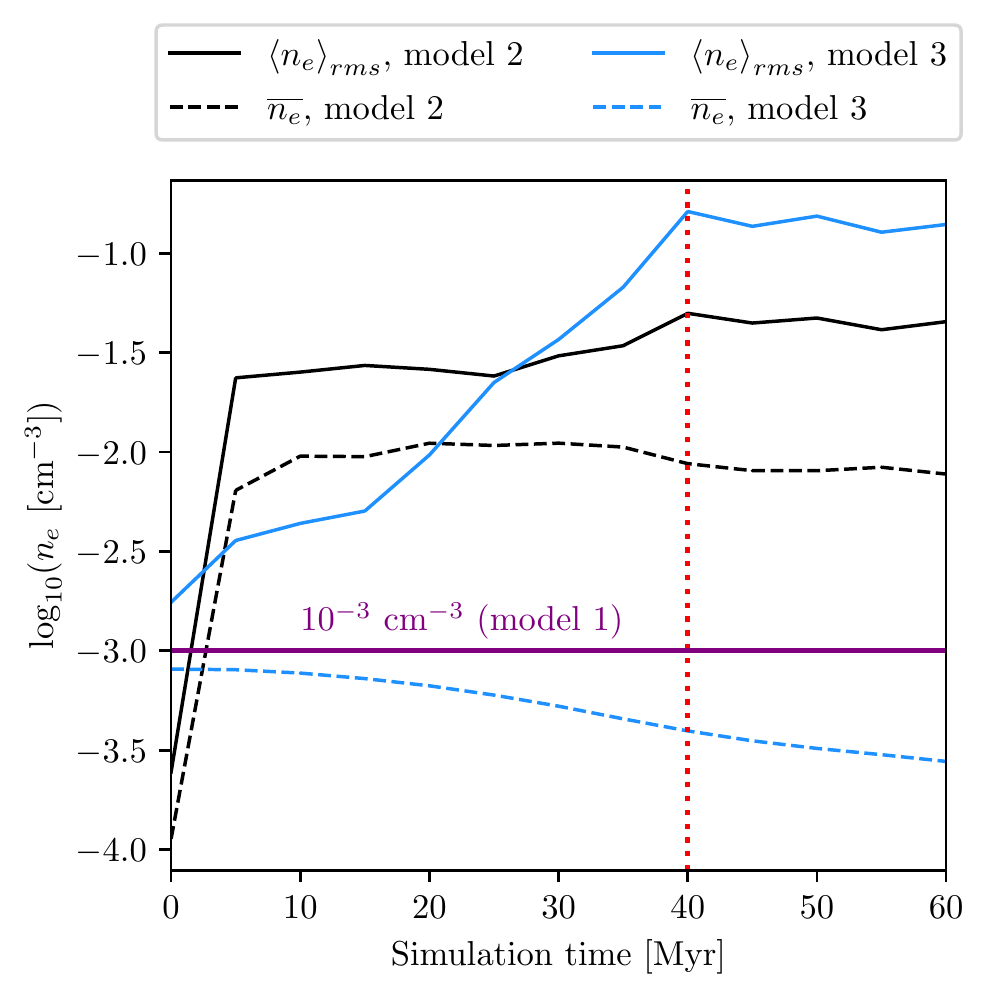}
    \caption{Time evolution of the volume-weighted average and rms of the thermal electron density $n_e$ over the whole 
    simulation domain, for $B_0 = 3$~$\mu$G. 
    The red dotted line corresponds to 40 Myr.
    This time is characteristic of the high intensity peak in the RM profiles for $n_e$ models 2 and 3 presented in Fig.~\ref{fig:rm_maps}.}
    \label{fig:ne_evolution}
\end{center}
\end{figure}

Figure~\ref{fig:ne_evolution} 
shows the evolution of the electron density for $n_e$ models 2 and 3. 
Both the rms and the average of $n_e$ model 2 are approximately constant over time, as is the average of $n_e$ model 3. 
In contrast, the rms for $n_e$ model 3 
increases by more than an order of magnitude
between about $20$ and $40$ Myr, and then 
becomes constant. 
This is directly linked to the fact
that $n_e$ model 3 takes into account all chemical species in the calculation of the electron density. 
Indeed, the zones of overdensity 
(due to shock waves from supernova explosions) 
tend to raise the rms value, the average value being unchanged or even very little influenced because these areas of high density are limited to small regions of space compared to the size of the simulation
box.

\subsection{Resulting rotation measure evolution in the simulations}
\label{subsec:FRM_results}

\begin{figure*}
\begin{center}
	\includegraphics[width=0.9\textwidth]{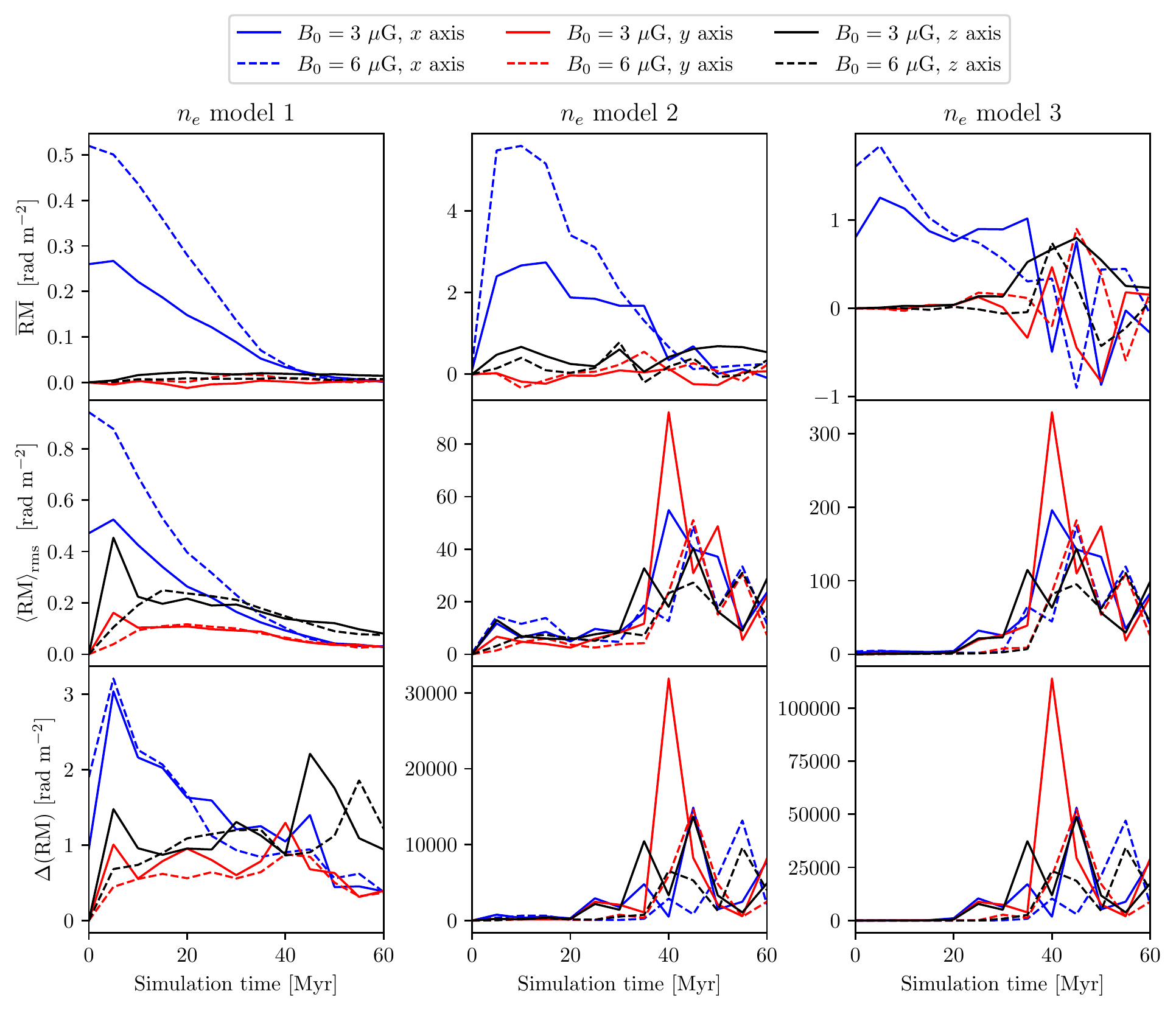}
    \caption{
    Time evolution of the RM for the three different $n_e$ models, calculated along each Cartesian axis, for $B_0 = 3$ (solid lines) and 6 $\mu$G (dotted lines). In particular, the plane average $\overline{\text{RM}}$ (top panels), the rms value $\left< \text{RM} \right>_{\mathrm{rms}}$ (middle panels), and the $\Delta$ estimator (equation \ref{eq:Delta_estimator}, bottom panels) are presented.}
    
    \label{fig:rm_maps}
\end{center}
\end{figure*}

\begin{figure*}
\begin{center}	
    \includegraphics[width=0.9\textwidth]{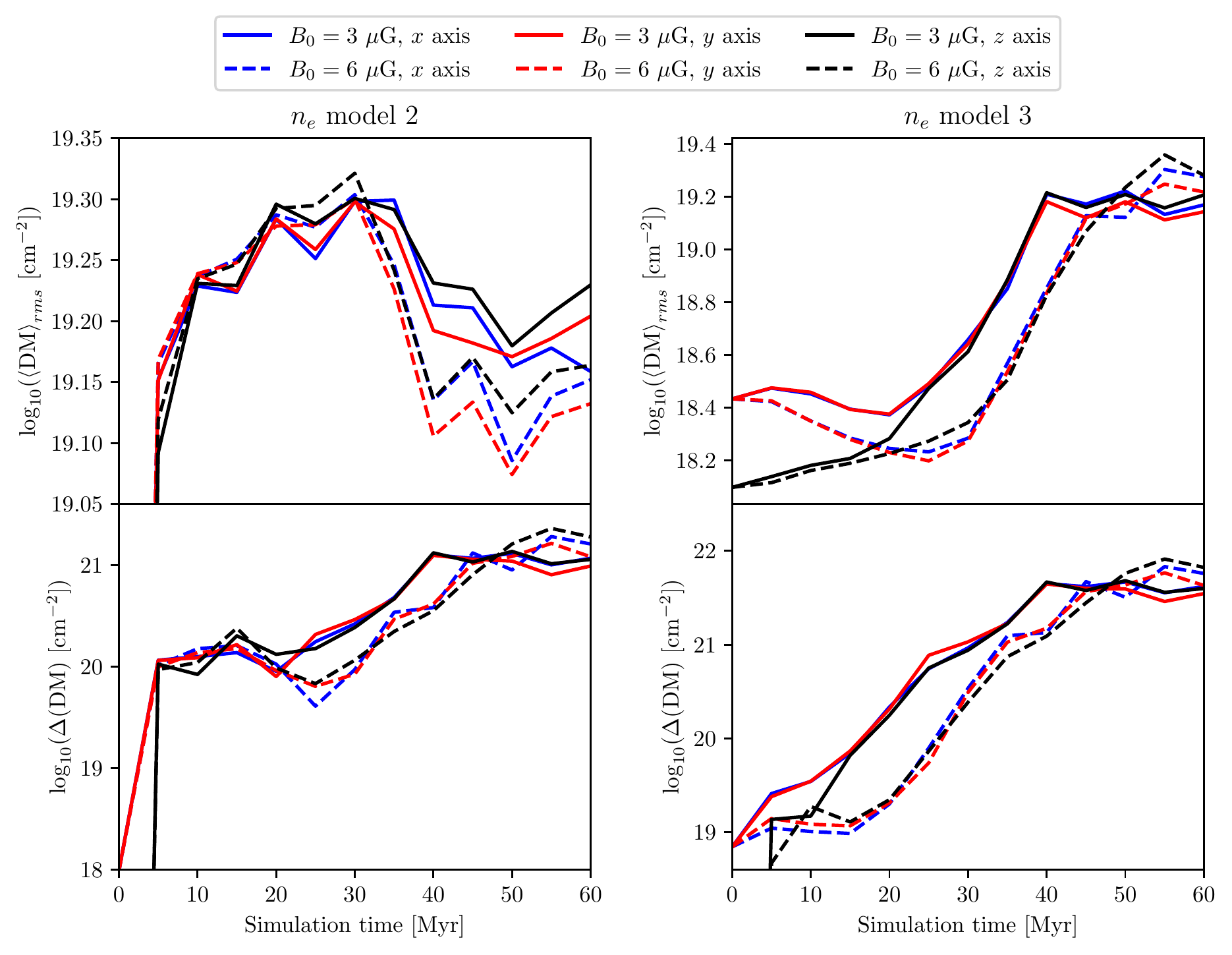}
    \caption{Dispersion measure profiles  for $n_e$ models 2 (left panels) and 3 (right panels) 
    with $f =10 ^{-3}$. 
    In particular, we calculate the rms value $\left< \text{DM}\right>_{\mathrm{rms}}$  
    (upper panels), and the estimator $\Delta$ given by Eq.~\eqref{eq:Delta_estimator}, 
    applied to the dispersion measure (lower panels).
    }
    \label{fig:dm_maps}
\end{center}
\end{figure*}

In Fig.~\ref{fig:rm_maps} we show the time evolution of
$\overline{\mathrm{RM}}$, $\left< \mathrm{RM} \right>_{\mathrm{rms}}$, and $\Delta(\mathrm{RM})$
in the three rows from top to bottom. 
The three $n_e$ models are shown from left to right for the two  
simulations with $B_0 = 3$ and 6 $\mu$G, respectively.

For the $n_e$ model 1 (left panels), it is not surprising to 
see that the mean and rms values of the RM evolve similarly to the magnetic field 
presented in Fig.~\ref{fig:RM_magfields_evolution}, because
this model uses a constant electron density (fixed at $n_e = 10^{-3}$ cm$^{-3}$). 
The $\Delta$ estimator decreases during the entire simulation time
(except for a small peak during the earliest times).
Along the $y$ and $z$-axes, $\Delta(\mathrm{RM})$ 
is almost constant during the entire simulation, with an 
average value around $\Delta(\mathrm{RM}) \simeq 1.5-2~\mathrm{rad}~\mathrm{m}^{-2}$. 
Along the $x$ axis, it constantly decreases over time, with a 
maximum value of $\Delta(\mathrm{RM}) \simeq 3~\mathrm{rad}~\mathrm{m}^{-2}$ at 5 Myr. 
This is mainly due to the fact that the magnetic field is initialized 
along the $x$ axis (decreasing exponential profile above and below the galactic plane), 
and it progressively spreads out in the simulation domain, loosing the initial 
disc-like shape, causing less variations between the average value of the magnetic 
field in the simulation box and its maximum value.

\begin{figure*}
\begin{center}
	\includegraphics[width=0.9\textwidth]{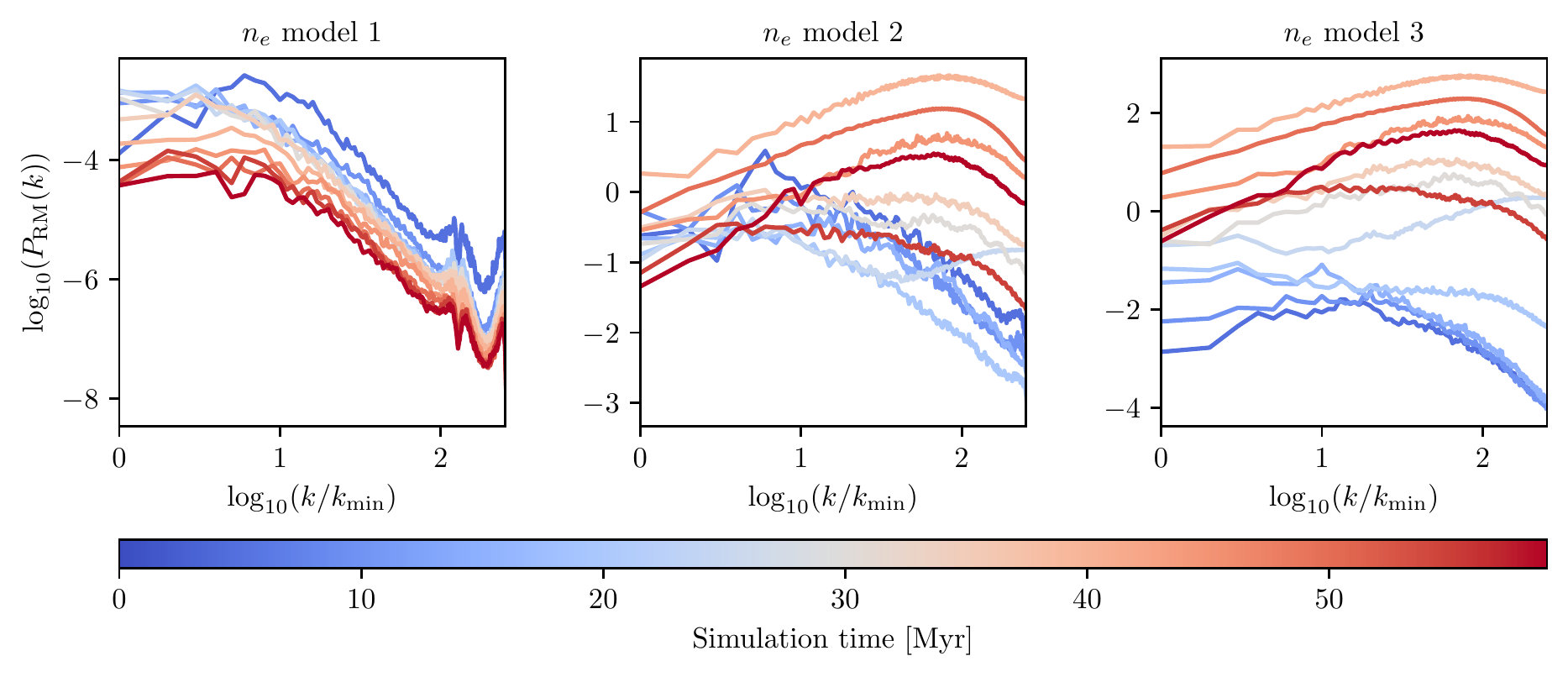}
    \caption{Power spectra of Faraday rotation maps of the SILCC simulation for $B_0 = 3$ $\mu$G, along the $y$ axis.
    The colors indicate the time of the spectra as given in the colorbar.  
    The analysis is shown for $n_e$ models 1 (left panel), 2 (middle panel), 
    and 3 (right panel).
    }
    \label{fig:rm_maps_spectrum}
\end{center}
\end{figure*}

The $n_e$ model 2 shows a different evolution. 
The value of $\overline{\mathrm{RM}}$ along the $y$ and $z$ axes is
approximately zero during the entire simulation time, which may indicate the isotropic 
character of the components of the magnetic field reached in particular 
because of the gas mixing. 
On the other hand, along the $x$ axis, $\overline{\mathrm{RM}}$ 
decreases over time to reach approximately zero at the end of the simulation, 
which is also an effect of the gas mixing.
A different evolution is observed for the rms value.
A peak (of approximately 60--80 rad m$^{-2}$, 
depending on the axis 
and the initial value $B_0$ considered) appears around 40 Myr. 
The presence of high-density zones of ionized species seems to be the 
origin of such a peak. 
Indeed, the curves $\Delta(\mathrm{RM})$ present the same kind of evolution, 
reaching more than $3\cdot10^4$ rad m$^{-2}$ along the $y$ axis. 
Furthermore, the evolution of $\Delta(\mathrm{DM})$
in Fig.~\ref{fig:dm_maps} shows an increase up to 30--40 Myr, 
reaching approximately 
$1.5\times10^{19}$~cm$^{-2}$ 
at the end of the simulation. 
Furthermore, $\Delta(\mathrm{DM})$ also increases during the whole simulation, 
supporting our hypothesis of the presence of 
high-density zones of ionized gas.
Finally, looking at the RM spectra shown in Fig.~\ref{fig:rm_maps_spectrum} 
(only along the $y$ axis), it appears that the peaks of the spectra are 
increasingly shifted towards larger wave numbers (i.e.\ shorter 
characteristic length scales). 
Overall, the typical RM values (plane average and rms) of $n_e$ model 2 
are about 10 to 20 times higher than those of $n_e$ model 1.

\begin{figure*}
\begin{center}
	\includegraphics[width=\textwidth]{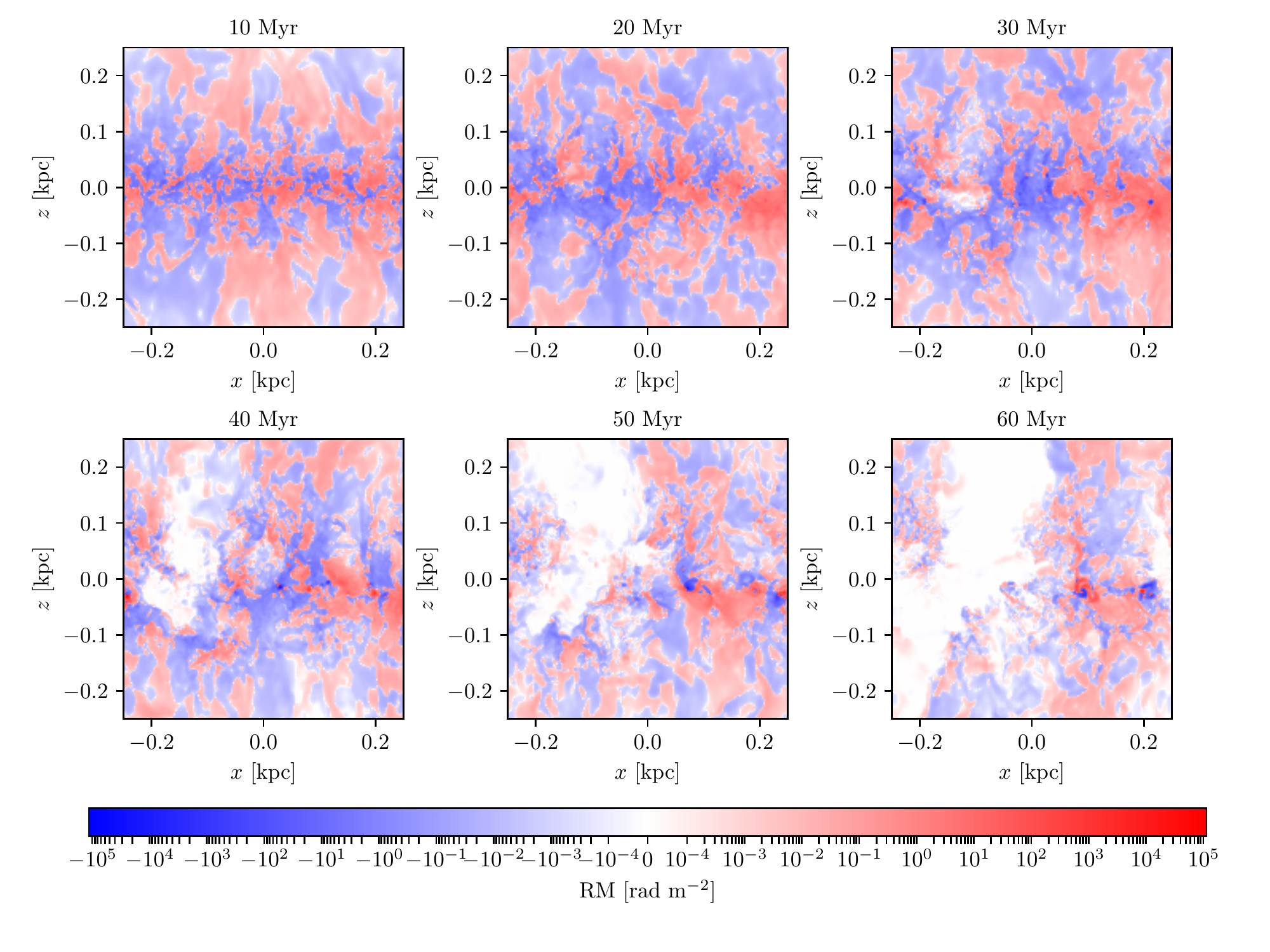}
    \caption{RM maps at different simulation times. 
    In this example, we present the RM
    along the $y$ axis, for $n_e$ model 3, and for
    $B_0 = 3$~$\mu$G.
     Note that we use a symmetrical logarithmic scaling in both the positive 
     and negative directions from the origin. 
     To avoid infinities around zero, a linear approximation 
     in the range from $-10^{-4}$ to $10^{-4}$ rad~m$^{-2}$ is adopted. 
    }
    \label{fig:rm_maps_example}
\end{center}
\end{figure*}

As mentioned before, in $n_e$ model 3, we have chosen to present only 
RM and DM results for the case $f = 10^{-3}$. 
This value was introduced in a \textit{ad hoc} manner in order 
to reproduce typical RM values that correspond
approximately to those observed in the Milky Way 
(e.g. \citealt{taylor_image_sky}). 
Changing $f$ would only shift the curves vertically, 
and therefore would not alter the conclusions. 
Figure~\ref{fig:rm_maps} shows that this $n_e$ model 
produces again different RM curves from the other two 
models discussed above. 
The evolution of $\langle\mathrm{RM}\rangle_{\mathrm{rms}}$ 
is extremely similar in both $n_e$ models 2 and 3, 
except that the peaks observed at 40 Myr reach almost 300 rad m$^{-2}$ 
(approximately four times the value observed in model 2). 
This may be due to the fact that the electron density 
is proportional to the total density of the gas, contrary to 
the $n_e$ model 2 which only takes into account the ionised species 
(C$^+$ and H$^+$).
The average value of the rotation measure has stronger variations than 
the model 2 after 40 Myr, which can be attributed to faster gas dispersion 
caused by stronger values of the magnetic field.  
Thus, the conclusions drawn for model 2 can partly be applied 
here as well: the dense gas that is compressed by supernovae 
and gravity can strongly contribute to $n_e$ in this model. 
Note that the presence of such zones with dense gas can be seen in 
Fig.~\ref{fig:rm_maps_example}, 
which shows RM maps for 
different simulation times along the $y$ axis.
Figure~\ref{fig:rm_maps_spectrum} shows the evolution of the 
RM power spectra for $n_e$ models 1, 2, and 3. 
In $n_e$ model 1, we observe that the spectra are 
peaked at the lowest wave numbers, 
which implies that the typical length-scale of RM patches is 
approximately the size of the simulation box.
This is not surprising given that the thermal electron density 
is assumed to be constant. 
We also observe that the peak of the spectrum is progressively 
shifted towards higher wave numbers, 
due 
to the restructuring of the magnetic field. On the contrary, 
$n_e$ models 2 and 3 show the opposite trend; at early times, 
the spectra resulting from $n_e$ model 2 have peaks around 
$\log_{10}(k/k_\mathrm{min}) \simeq 0.8$ which are progressively 
shifting towards the largest wave numbers, while for $n_e$ model 3, 
this trend is even more prominent 
(It appears nevertheless that the 
initial peaks of $n_e$ model 3 are more around 
$\log_{10}(k/k_\mathrm{min}) \simeq 1.2-1.5$, which can be simply 
explained by the fact that all of the chemical species are included 
in the calculation of $n_e$ in this model, and that the density of 
electrons can be more important in certain regions other than in $n_e$ model 2). 
The most probable explanation for such an observation is that the 
formation of regions with high electron density is formed by the 
SN-driven turbulence. 
Despite the fact that these power spectra only allow us to extract 
qualitative characteristics of the distribution of RM values, they are, 
in our case, a useful source of information allowing us to support our 
hypothesis that areas of high electron density are responsible for the 
different patterns observed between the thermal electron models.

\subsection{Impact of the initial large-scale magnetic field on the Faraday
signals}

It should be noted that the initial conditions for the 
magnetic field in the SILCC simulation are simplified. 
It is initialized along the $x$ axis, that does not take 
into account the complexity of the magnetic field that 
could be observed in spiral galaxies. 
However, we argue that this condition should not 
affect our conclusions.

Observations of face-on spiral galaxies show the presence 
of a large-scale magnetic fields with a component following the direction of spiral arms (e.g. \citealt{Beck_spiral_galaxy_magfield}). 
Therefore, considering a large-scale magnetic field along the $x$-direction (the galactic plane lies in the $xy$-plane) should be a good approximation. 
Also, the simulation box is a cube of 500 pc edges, 
which is quite small compared to the overall size of the Milky Way (whose radius is though to be of the order of 20 kpc). 
Therefore any geometrical variations of the field 
should not be relevant at this scale. 
However, the observations reveal that the large-scale magnetic 
field is about ten times smaller than the random field, which is not reflected in the initial conditions.

However, the magnetic field evolves dynamically throughout the simulation
and the field configuration is more realistic after 
approximately $20$~Myr. 
Therefore, our conclusions are only based on the data at times $\gtrsim20$~Myr. 
On the contrary, above approximately 45-50 Myr, a non negligible amount of gas has left the box, and the dynamics has altered the gas structure significantly.


\section{Synchrotron emission}
\label{sec:SYNCH_emission}

\subsection{Theory and methods}
\label{sec:SYNCH_methodology}

\subsubsection{Diffusion-loss equation for the cosmic ray power spectrum}
\label{subsubsec:SYNCH_basic_equations_and_modelling}
The synchrotron luminosity is a second observable of the magnetic field in the ISM. 
Synchrotron emission is produced by cosmic ray 
electrons gyrating around magnetic fields.
Since cosmic rays are not included in the DR6 data release of the 
SILCC simulations, we employ a semi-analytical model for the power 
spectrum of such electrons.
We follow the approach adopted by \cite{schober_galactic_synchrotron_emission}, but see also \citet{WerhahnEtAl2021a,WerhahnEtAl2021b,maria_paper_cosmic_rays} for a similar approach on galactic scales.

Our model includes two different 
populations of CR electrons. 
The primary population, which originates from supernovae remnants
(meaning that electrons gain energy through Fermi acceleration 
processes) and the secondary electrons, which are created by 
proton-decay pion production. 
The general expression for the energy injection spectrum of a cosmic ray species $\alpha$ is a power-law \citep{bell_shockwaves_fermi_1} of the form
\begin{equation}\label{power_spectrum_power_law}
   Q_{\alpha}(E) = Q_{\alpha,0} \left(\frac{E}{m_{\alpha}c^2}\right)^{-\chi},
\end{equation}

where $Q_{\alpha,0}$ is the normalization factor, 
$E$ is the energy of species $\alpha$ (rest mass plus kinetic energy), and $\chi$ is the spectral index.
Analytical models of cosmic ray acceleration 
predict a dependence of $\chi$ on the compression factor and characteristic 
velocities of the shock \citep{bell_shockwaves_fermi_1}.
For typical shock parameters, these analytical models result in $\chi=2.0-2.5$, 
but more detailed models of supernova shock fronts yield $\chi=2.1-2.3$
\citep{BogdanVolk1983, Lacki_2012_supernova_shocks}. 
We choose a value of $\chi = 2.2$ for our study.
The evolution of the spectrum for a cosmic ray species
$\alpha$, $N_{\alpha}(E)$, is governed by the 
diffusion-loss equation \citep[see e.g.][]{torres2004_CR}
\begin{equation}\label{diffusion_loss_equation}
\frac{\partial N_{\alpha}(E)}{\partial t} = Q_{\alpha}(E) + \frac{\mathrm{d}}{\mathrm{d}E} 
\left[ b(E) N_{\alpha}(E)\right]
- \frac{N_{\alpha}(E)}{\tau_{\alpha}(E)}+D\nabla^2N_{\alpha}(E),
\end{equation}
where $b(E) = -\mathrm{d}E/\mathrm{d}t$ is the cooling rate, 
$\tau_{\alpha}(E)$ is the timescale for catastrophic losses, and $D$ is a spatial diffusion coefficient.

\subsubsection{Primary CR electrons} 
In order to derive the steady state spectrum of primary electrons, 
Eq.~\eqref{diffusion_loss_equation} 
can be simplified by making 
the same assumptions as proposed in \citet{pionic_secondary_energy_losses}.
For a steady state, the time derivative in Eq.~\eqref{diffusion_loss_equation} vanishes. 
Additionally, one assumes spatial homogeneity of the ISM, and 
that catastrophic losses are negligible. 
The cooling rate is assumed to be of the form 
$b(E) \simeq E/\tau_e(E)$, where $\tau_e(E)$ represents the typical 
cooling timescale of the CR electrons, 
involving various processes that 
will be described below. 
Therefore, 
Eq.~\eqref{diffusion_loss_equation} simply becomes:
\begin{equation}
    Q_e(E) = -\frac{d}{dE}\left( \frac{EN_e(E)}{\tau_e(E)}\right).
\end{equation}
After integrating the latter with respect to $E$, we have
\begin{equation}\label{initial_CRe_power_spectrum}
    N_e(E) = \frac{Q_e(E)\tau_e(E)}{\chi-1}.
\end{equation}

\subsubsection{Secondary CR electrons and CR protons}
In Eq.~\eqref{initial_CRe_power_spectrum}, the injection
spectrum $Q_e(E)$ of CR electrons takes into account the
contribution of both primary and secondary electrons, namely 
$Q_e(E) = Q_{e,\mathrm{prim}}+ Q_{e, \mathrm{sec}}$. 
However, 
using gamma-ray observations of M82 and
NGC253, \citet{Lacki_et_al_M82_ngc253}, 
\citet{pionic_secondary_energy_losses} 
estimated that the value of the ratio 
$f_{\mathrm{sec}}\equiv Q_{e, \mathrm{sec}}/(Q_{e, \mathrm{prim}}+Q_{e, \mathrm{sec}})$ 
to be approximately 0.6-0.8. 
In this paper, we adopt $f_{\mathrm{sec}}=0.7$. 
Therefore, we have $Q_e = Q_{e, \mathrm{sec}}/f_{\mathrm{sec}}$. 
Following again the steps of \citet{pionic_secondary_energy_losses}, we can relate 
$Q_{e, \mathrm{sec}}$ to the power spectrum of CR protons $Q_p$. 
Note that the CR protons mainly suffer catastrophic losses via pions production.
The typical life time of CR protons is $\tau_{\pi}f_{\pi}$ with $\tau_\pi$ being the timescale of pion production. 
We set $f_{\pi} = 0.4$ \citep{pionic_secondary_energy_losses}.
Therefore, determining $Q_p$ is sufficient to find 
the full expression of 
the CR electrons. 
The injection spectrum of CR protons is estimated as follows. 
We make the 
hypothesis that the protons only 
originate from supernovae explosions. 
Assuming a power-law expression of the form 
\eqref{power_spectrum_power_law}, and connecting $Q_p$ to 
the supernova rate (see \citealt{schober_galactic_synchrotron_emission}), we can write the expression
\begin{equation}\label{equ_SN_rate}
    \int_{m_pc^2}^{\infty} Q_{p,0}\left( 
    \frac{E}{m_pc^2}\right)^{-\chi}E~\mathrm{d}E 
    = \xi \dot{n}_{\mathrm{SN}}E_{\mathrm{SN}},
\end{equation}
where $\xi$ is the amount of supernovae energy transferred into 
energy of CR protons, 
$\dot{n}_{\mathrm{SN}} \equiv \dot{N}_{\mathrm{SN}}/\mathrm{d}V$ is the 
supernova rate density with $\dot{N}_{\mathrm{SN}}$ being the supernova rate per grid cell 
and $\mathrm{d}V$ the volume of a grid cell, and $E_{\mathrm{SN}}$ is the total energy released by a 
single supernova. 
We assume $\xi = 0.1$ and $E_{\mathrm{SN}} = 10^{51}$ erg. Integrating the latter equation and rearranging all the terms yields
\begin{equation}\label{protons_injection_spectrum}
    Q_{p,0} = \frac{1}{\mathrm{d}V}\frac{(\chi-2)\xi
    \dot{N}_{\mathrm{SN}}E_{\mathrm{SN}}}{(m_pc^2)^2}.
\end{equation}
Note that $\dot{N}_{\mathrm{SN}}$ is derived according to how the 
probability of supernovae is implemented 
in the SILCC simulations (see \citealt{SILCC_project_molecularclouds}), and we provide a detailed
calculation of this term in Appendix~\ref{appendix:numerical_SN_rate}.
The factor $1/\mathrm{d}V$ ensures to have the right units for $Q_{p,0}$.
Finally, putting Eqns.~\eqref{power_spectrum_power_law}, \eqref{initial_CRe_power_spectrum},
and \eqref{protons_injection_spectrum} together, we obtain
\begin{equation}\label{CR_e_power_spectrum}
\begin{split}
&N_e(E) =\\ 
&\frac{1}{\mathrm{d}V}\frac{f_{\pi}}{f_{\mathrm{sec}}}\frac{20^{2-\chi}}{6}
\frac{(\chi-2)\xi \dot{n}_{\mathrm{SN}}E_{\mathrm{SN}}}{(m_pc^2)^2} \left( \frac{E}{m_pc^ 2} \right)^{-\chi}\tau_e(E).
\end{split}
\end{equation}

\subsubsection{CR electron cooling timescale}
\label{subsubsec:cooling_timescales}

In Eq.~(\ref{CR_e_power_spectrum}), the different cooling processes taken
into account in the expression of $\tau_e$ are ionization (ion),
bremsstrahlung (brems), inverse Compton scattering (IC), 
synchrotron emission (synch), and galactic winds (wind). 
Their expressions are given as follows:

\begin{eqnarray}
&\tau_{\mathrm{ion}}& = \frac{E/(m_ec^2)}{2.7c \sigma_t \left(6.85+0.5\ln\left(\frac{E}{m_ec^2}\right)\right)n_{\mathrm{neut}}},\label{ion_timescale}\\
&\tau_{\mathrm{brems}}& = 3.12\times10^{7}~\mathrm{yr} \left( \frac{n_{\mathrm{ion}}}{\mathrm{cm}^{-3}}\right)^{-1},\label{brems_timescale}\\
&\tau_{\mathrm{IC}}& = \frac{3 m_e c}{4 \sigma_t u_{\mathrm{ISRF}}\frac{E}{m_ec^2}},\label{IC_timescale}\\
&\tau_{\mathrm{synch}}& = \frac{3 m_e c}{4 \sigma_t u_\mathrm{mag}\frac{E}{m_ec^2}},\label{synch_timescale}\\
&\tau_{\mathrm{wind}}& = \frac{H}{v_{\mathrm{wind}}},\label{wind_timescale}
\end{eqnarray}
where $\sigma_t$ is the Thompson cross section, $n_{\mathrm{neut}}$
is the density of all neutral species implemented in the 
simulations (namely H, H$_2$, and CO), $n_{\mathrm{ion}}$ is 
the density of ionized species (H+ and C+), $n_{\mathrm{ISRF}}$ is
the energy density of the interstellar radiation field, 
$u_\mathrm{mag}$ is 
the energy density of the magnetic field, $H$ is the typical length 
scale of the simulation domain, and $v_{\mathrm{wind}}$ is the 
typical velocity of galactic winds. 
The ionisation formula comes 
from \citet{ion_cooling_process}, and the bremsstrahlung expression 
comes from \citet{brehm_cooling_process}. 
In $\tau_{\mathrm{wind}}$, we adopt $H = 300$ pc, which is the scale 
height at which the stellar component of the gas is initialized in the SILCC simulations. For 
the typical value of the galactic winds, we adopt $v_{\mathrm{wind}} = 
50$ km/s.

In order to derive the expression of the energy of the interstellar radiation field, we adopt the same approach as \cite{schober_galactic_synchrotron_emission}. 
We consider four main components: the cosmic microwave background (CMB), 
infrared (IR), an optical component (opt) and finally an ultraviolet
(UV) component  \citep[see also][]{WinnerEtAl2019,WinnerEtAl2020}.
If we assume that each of those components can be
described by a Planck curve with the corresponding temperature, 
then the total 
energy density of thermal interstellar radiation is
\begin{equation}\label{equ:SILCC_syhcnrotron_uISFR}
\begin{split}
u_{\text{ISFR}} &=  \int_{0}^{\infty} \left[ \sum_i f_i\frac{8\pi h}{15c^3h^3}\frac{\nu^3}{e^{\left( \frac{h \nu}{kT_i}\right)}-1}\right]~\mathrm{d}\nu\\
&= \frac{8 \pi^5 k^4}{15 c^3 h^3}\sum_i f_iT_i^4
\end{split}
\end{equation}
with $i \in \{\mathrm{UV}, \mathrm{opt}, \mathrm{IR}, \mathrm{CMB}\}$.
The coefficients $f_i$ and the temperatures $T_i$ were estimated by \cite{ISRF_fi_factors} and \cite{ISRF_fi_factors_2}, and are summarized 
in Tab.~\ref{tab:SILCC_syhcnotron_fi_Ti}.

\begin{table}
	\centering
	\caption{Weight factors and temperature of each thermal component considered in formula \eqref{equ:SILCC_syhcnrotron_uISFR}. 
 	Those factors were calculated by \citet{ISRF_fi_factors} and \citet{ISRF_fi_factors_2}.}
	\label{tab:SILCC_syhcnotron_fi_Ti}
	\begin{tabular}{lcc} 
		\hline
		Process & $f_i$ & $T_i~[K]$ \\
		\hline
        UV & $8.4 \cdot 10^{-17}$ & $1.8 \cdot 10^8$\\
        Optical & $8.9 \cdot 10^{-13}$  & $3.5\cdot 10^{3}$\\
        IR & $1.3 \cdot 10^{-5}$  & 41\\
        CMB & 1 & $2.73$\\
		\hline
	\end{tabular}
\end{table}

\begin{figure*}
    \centering
    \includegraphics[width = \textwidth]{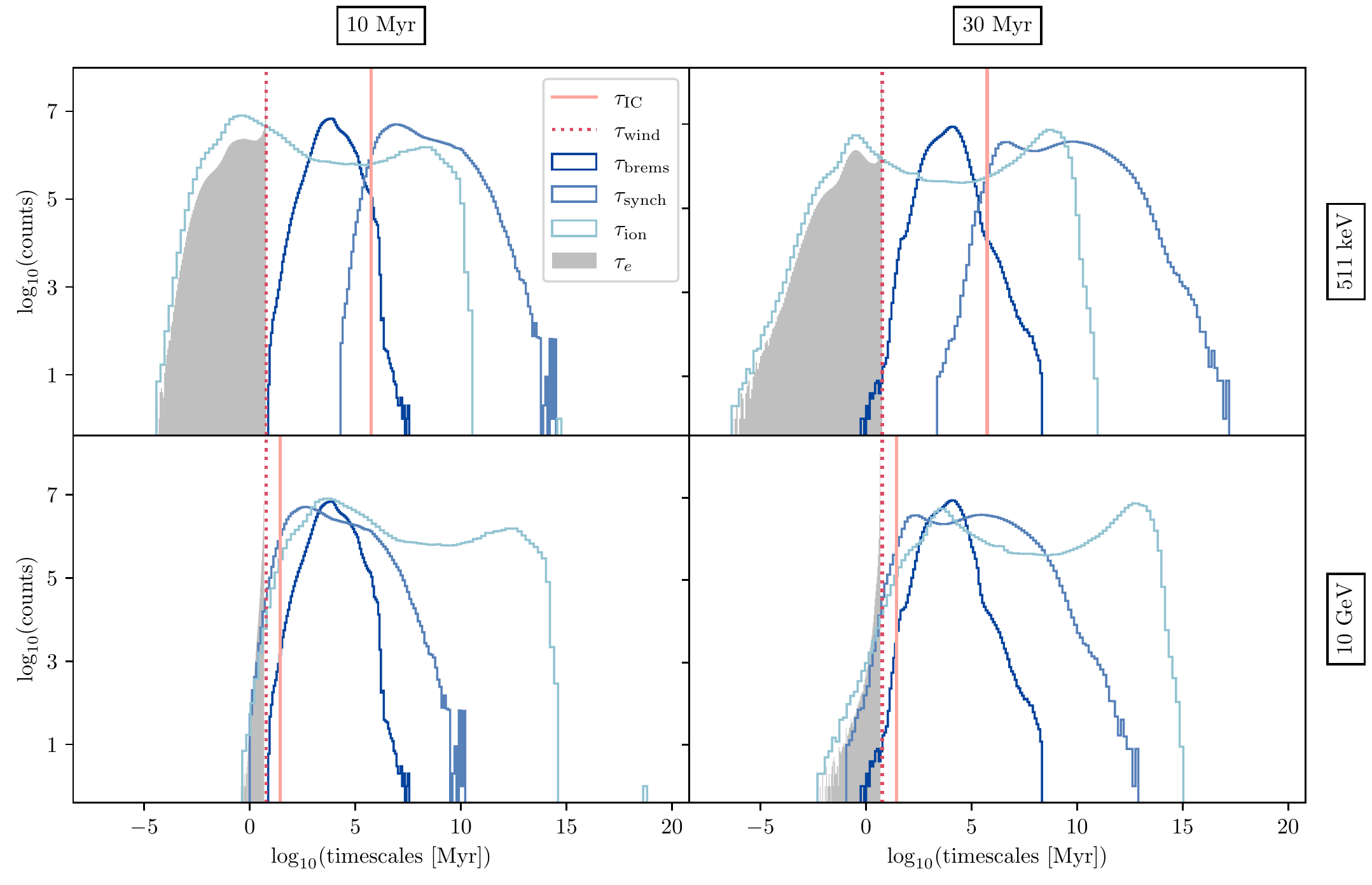}
    \caption{Histograms of the different cooling timescales involved in the expression of 
    Eq.~\eqref{eq:tau} entering the diffusion-loss 
    Eq.~\eqref{diffusion_loss_equation} 
    (for the whole three-dimensional 
    simulation box), at 10 and 30 Myr, for 511 keV (electron mass energy) and 10 GeV, and for 
    $B_0 = 3$~$\mu$G.
}
    \label{fig:3d_cooling_timescales_b03}
\end{figure*}

Altogether, the general expression of the cooling timescale is given as:
\begin{equation}\label{eq:tau}
    \tau_e = \left( \tau_{\text{ion}}^{-1}
    +\tau_{\text{brems}}^{-1}
    +\tau_{\text{IC}}^{-1}+\tau_{\text{synch}}^{-1}+\tau_{\text{wind}}^{-1}\right)^{-1}.
\end{equation}
Figure~\ref{fig:3d_cooling_timescales_b03} shows histograms of 
the different cooling timescales given by 
Eqs.~\eqref{ion_timescale}--\eqref{wind_timescale} 
in all grid cells, at $10$ and $30$ 
Myr and for cosmic ray energies of $E=511$~keV and $10$~GeV, for 
the run with $B_0 = 3$~$\mu$G. 
Clearly, it appears that at 10 GeV, in most cells 
$\tau_e$ is determined by $\tau_\mathrm{wind}$,
which, in our model, does not depend on the energy of the cosmic rays and is constant 
in space and time. 
For lower energies, 
the dominant contribution to $\tau_e$ 
is ionisation. 
Furthermore, 
bremsstrahlung becomes 
relevant for later times, given the cooling timescale associated with
this process depends on the density of ionized species, that are 
constantly created by the supernovae explosions. 
At $E = 10$~GeV, 
given its dependency on the energy, the 
synchrotron emission starts to play a role (although not major) in the 
values of $\tau_e$. 
The inverse Compton scattering has not yet crossed the value of 
$\tau_{\mathrm{wind}}$, but it will likely take over the galactic 
winds as an upper limit of $\tau_{e}$. 
As an example, \cite{maria_paper_cosmic_rays} 
modelled various cooling 
timescales for disc-like galaxies, implementing non-radiative (hadronic
and Coulomb interactions of CRs with the interstellar medium) and 
radiative (synchrotron, inverse Compton, and 
bremsstrahlung) processes. 
They showed an example  of typical cooling timescales value at 10 GeV 
(see Fig.~1 in \citealt{maria_paper_cosmic_rays}). 
Each of these processes rarely has a characteristic timescale longer 
than approximately $10^3$ Myr. 
In our study, although those values seem to match the 
ones of our implemented processes, numerous grid cells reach extreme 
values that could go up to approximately $10^{12}$ Myr in the case of 
the synchrotron emission. At first glance, this is clearly due to the 
gas distribution of the simulations, which in our case create regions 
with extreme values of neutral (or ionized) gas densities. 
Our results show that the gas distribution in the ISM along with the choice of
cooling processes governing the diffusion-loss equation of the cosmic rays are two major components that could impact the power spectrum of CR electrons to a large extent. 
Note that no major differences are observed with the case 
$B_0 = 6$~$\mu$G 
(see Fig.~\ref{fig:3d_cooling_timescales_b06} in the appendix), so all our analysis can apply to this case without any loss of generality.

\subsubsection{Synchrotron luminosity}
\label{subsubsec:SYNCH_basic_equations}
Using the work of \cite{blumenthal_radiation}, the expression of the synchrotron power spectrum for a distribution of electrons is given by:

\begin{equation}
\label{equ:SILCC_synch_lum_power_formula}
\mathcal{L}(\nu) = \int_{m_ec^2}^{\infty}P_e(\nu, E)N_e(E)~\mathrm{d}E
 \times\int N(\alpha)(\sin \alpha)^{\frac{\chi+1}{2}}~\mathrm{d}\Omega 
\end{equation} 
where
\begin{equation}
  P_e(\nu, E) = \frac{\sqrt{3}e^3B}{m_ec^2}\frac{ \nu}{\nu_c} \int_{\nu/\nu_c}^{\infty} \mathcal{K}_{5/3}(x)~\mathrm{d}x
\end{equation}
with $\mathcal{K}_{5/3}$ being the Bessel function of the second kind
with parameter $5/3$ is the power spectrum of a single electron, and 
where 
\begin{equation}
    \nu_c \equiv \frac{3E^2eB}{4\pi m_e^3c^5}.
\end{equation}
The integral over the pitch angle $\Omega$ is
roughly equal to 8.9, for the specific value $\chi = 2.2$. 
Note that \cite{schober_galactic_synchrotron_emission} also added the contribution 
of free-free electron absorption/emission. 
However, with the typical values of the physical parameters we consider in our models, this contribution is negligible, and therefore not considered in our model.

\subsubsection{Data treatment}
\label{subsubsec:SYNCH_data_treatment}
As mentioned in Sec.~\ref{subsubsec:FRM_data_treatment}, the resolution of the simulations is dynamically and locally adapted with respect to the gas density. 
In particular, $N = 512^3$ corresponds to the resolution of the most refined regions in the simulations. 
We adopt the same strategy described in 
Sec.~\ref{subsubsec:FRM_data_treatment}, and map all the data onto a 
uniform grid of $N = 512^3$ grid cells, except for the data 
corresponding to $t = 0$ Myr, for which the resolution
is fixed at $N = 128^3$. 
In particular, this means that less refined regions are simply volume-weight averaged. 
In order to work on two-dimensional maps, we reduce the three-dimensional data in the following way. 
For a given physical quantity $\Phi$ (given as an output of the simulations), we calculate the weighted average
\begin{equation}\label{average_projection}
\langle \Phi \rangle(i,j) \equiv \frac{1}{L}\sum_{k=0}^{N_{\mathrm{grid}}} \Phi(i,k,j)\Delta l_k
\end{equation}
where $L = 500$ pc is the size of the simulation domain and $\Delta l_k \equiv L/N_{\mathrm{grid}}$ is the size of a grid cell. 
Given that our data are mapped onto a uniform grid, 
Eq.~\eqref{average_projection} simply reduces to the sum of 
$\Phi$ along a given axis. 
By convention, we perform our analysis along the $y$ axis, meaning 
that we work with the $xz$ plane, $z$ being the vertical coordinate from 
the disc plane.
We apply then Eq.~\eqref{average_projection} to the components 
of the magnetic field, and to gas fractions of the different chemical 
species involved in the dynamics of the simulations.

The synchrotron luminosity \eqref{equ:SILCC_synch_lum_power_formula} 
is directly calculated on averaged two-dimensional maps. 
In particular, we calculate the quantity
$L_{1.4\mathrm{GHz}} \equiv 1.4 \; \mathrm{GHz}~ \mathcal{L}(1.4 \; \mathrm{GHz})$ 
that will be presented in units of solar luminosity 
$L_{\odot} = 3.826 \cdot 10^{33}$~erg~s$^{-1}$.
For quantitative comparison, we create a one-dimensional profile for maps of a physical quantity of interest as follows. 
Once a projected map is created according to formula \eqref{average_projection}, 
we take the average value, using the same formula \eqref{equ:FRM_plane_formula} for the average of rotation measure maps, of each horizontal line (corresponding to a fixed coordinate above/below the galactic midplane). 
As an example, 
for calculating the vertical profile of the magnetic field amplitude, 
we project each magnetic field component
$B_x, B_y$, and $B_z$ along a given axis 
($y$ axis across the midplane by convention), 
we create the map of the projected magnetic field as 
$B = (B_x^2+B_y^2+B_z^2)^{1/2}$, and then we take the 
average value of each horizontal line. 
This process could be applied to any physical quantity 
that is given as an output of the simulations.

\subsection{Resulting synchrotron luminosity in the simulations}
\label{subsec:SYNCH_results}

\begin{figure*}
    \centering
    \includegraphics[width=\textwidth]{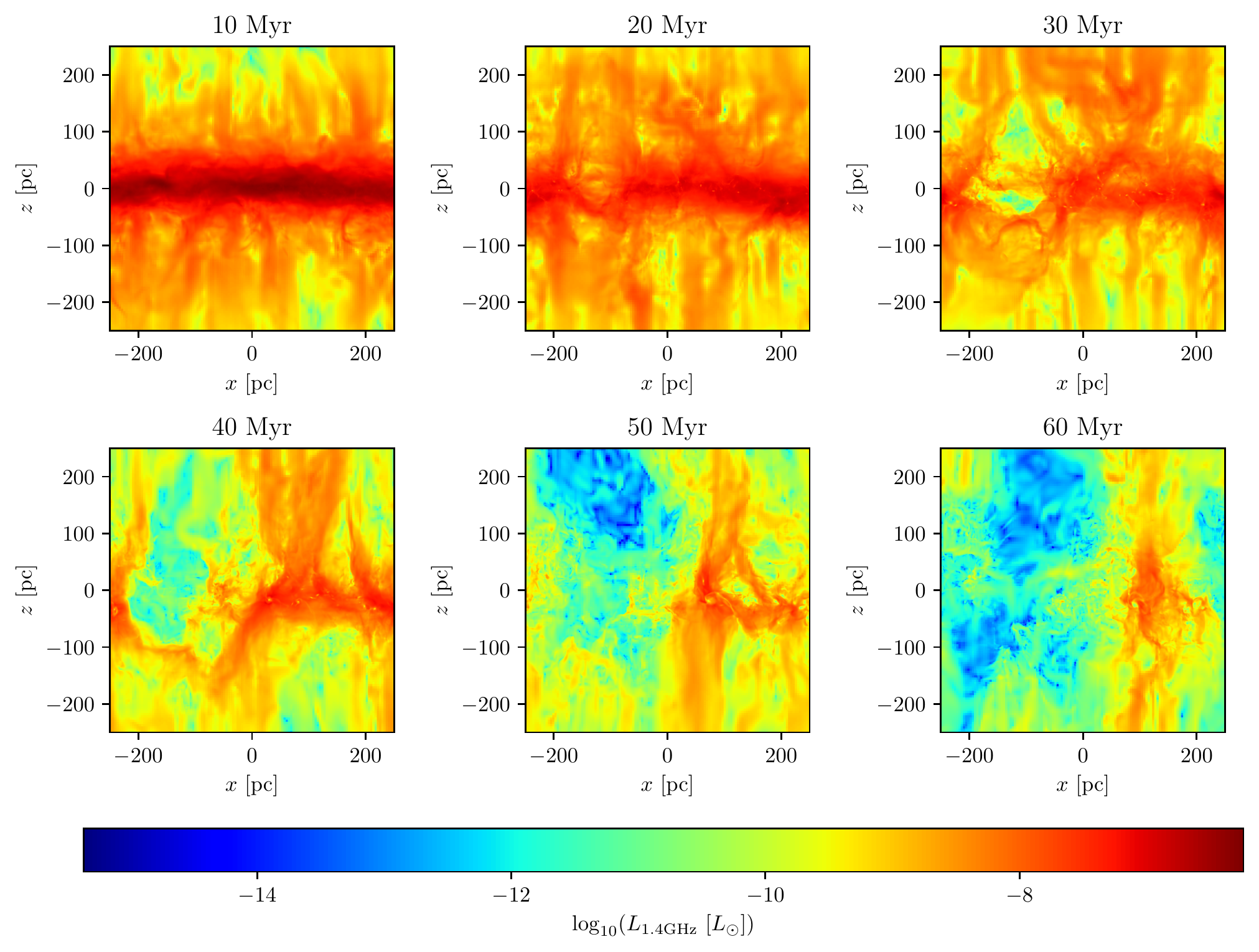}
    \caption{Synchrotron luminosity maps in the $xz$ plane for different simulation times. The initial disc-shaped gas structure can be observed up to about 
    $20-30~\mathrm{Myr}$.
    Thereafter, magnetic pressure and supernova explosions disperse the gas. 
    The synchrotron luminosity $L_{1.4\mathrm{GHz}}$ (expressed in units of the solar luminosity $L_\odot$) given in the color bar, decreases with time.
    This is mainly due to the fact that neutral hydrogen is progressively either being ionized, 
    transported out of the simulation domain, or condensed to small dense clouds.
    }
    \label{fig:SYNCH_luminosity_maps_example}
\end{figure*}
\begin{figure}
    \centering
    \includegraphics[width=\columnwidth]{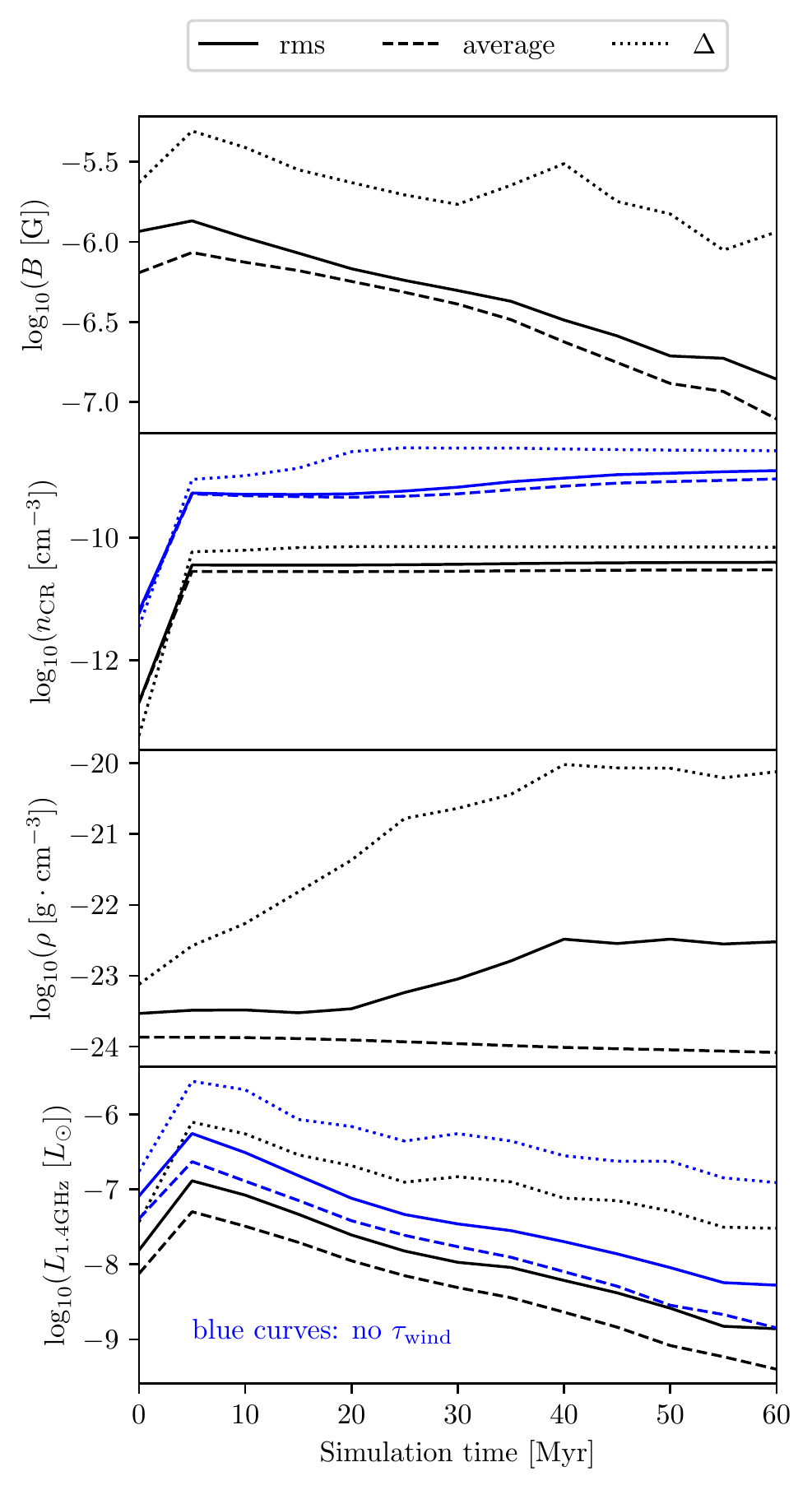}
    \caption{Starting from the top: time evolution evolution of three estimators (average, rms and $\Delta$) of the magnetic field, 
    of CR electron density given by Eq.~\eqref{n_CR}, of the total gas density, and of the synchrotron luminosity $L_{1.4\mathrm{GHz}}$, for $B_0 = 3$ $\mu$G. 
    The blue curves correspond to synchrotron luminosity calculated by omitting the contribution of $\tau_{\mathrm{wind}}$ in the expression~\eqref{CR_e_power_spectrum}.
    }
    \label{fig:B_intNe_rho_L}
\end{figure}

Figure~\ref{fig:SYNCH_luminosity_maps_example} shows an example of the
evolution of synchrotron luminosity maps, in the $xz$ plane, 
for the run with $B_0 = 3$ $\mu$G. 
We observe that the typical values of the
luminosity globally decrease with time. 
Additionally, the highest values seem to be correlated with the 
density of the gas. 
Indeed, it can be observed on the $10$ Myr map that most of the
synchrotron luminosity is concentrated in the center of the galactic 
mid-plane. On the other hand, we cannot exclude that the synchrotron 
luminosity is mostly influenced by the value of the magnetic field, 
which would be more in line with our intuition. 
A first indication is given in Fig.~\ref{fig:B_intNe_rho_L}, which shows the time evolution of one-dimensional profiles of the magnetic field, the cosmic ray number density given by 
\begin{equation}\label{n_CR}
  n_{\mathrm{CR}} \equiv \int_{m_ec^2}^{\infty}N_e(E)dE,
\end{equation}
where $N_e$ is given by Eq.~\eqref{CR_e_power_spectrum}, the gas total density, 
and the synchrotron luminosity. 
It becomes then more evident that the decrease of the synchrotron luminosity is rather connected to the evolution of the magnetic field than the other quantities. Indeed, it appears that the average gas density does not vary significantly in time, while
the $\Delta$ and rms values increase (which clearly 
indicates that denser regions are created by shock waves following 
supernova explosions).

\begin{figure}
    \centering
    \includegraphics[width=\columnwidth]{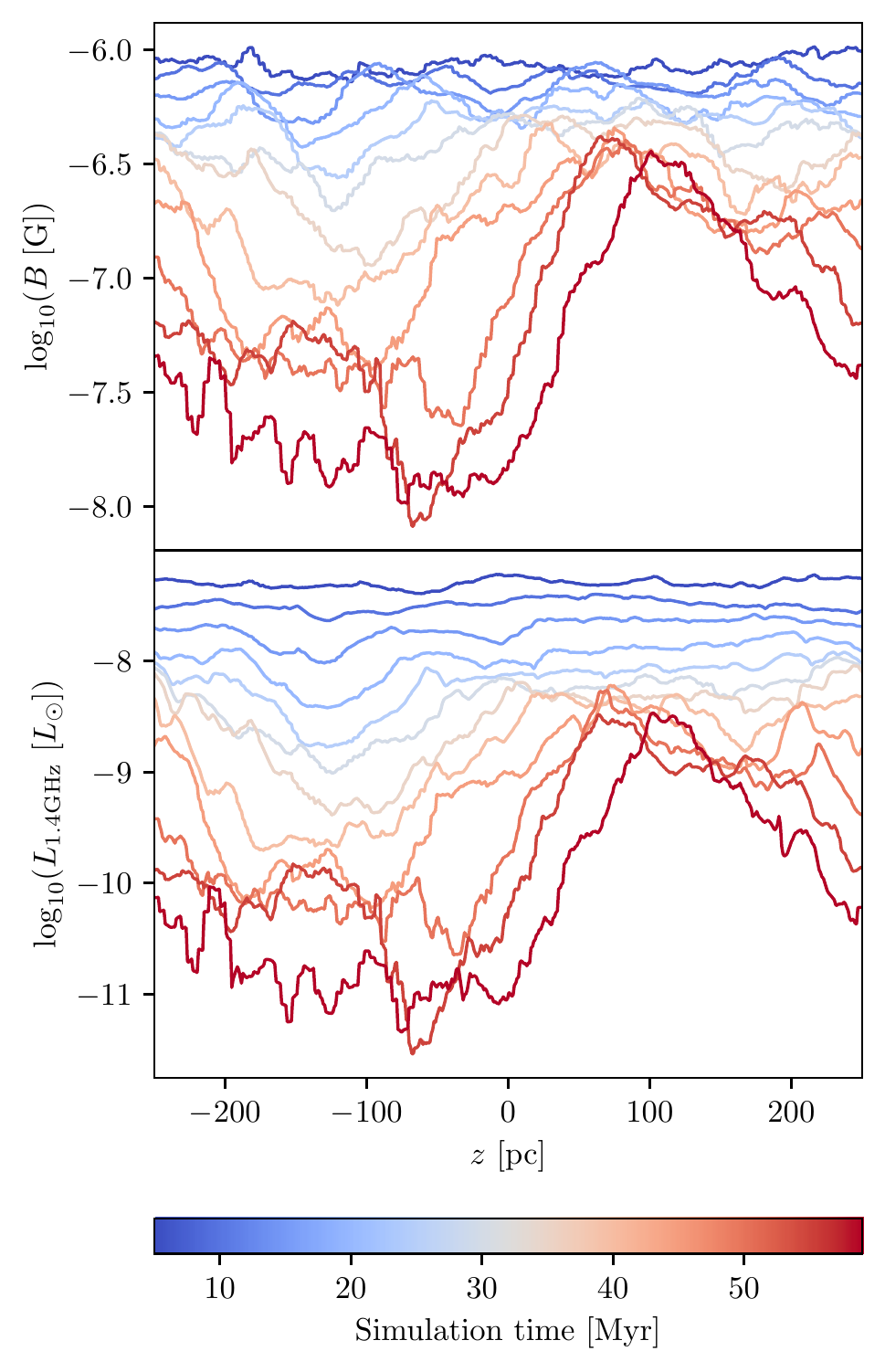}
    \caption{Vertical profiles of the volume-weighted projected  magnetic field
    (top), and synchrotron luminosity (bottom), along the $y$ axis, 
    for $B_0 = 3$ $\mu$G.}
    \label{fig:B_vs_synchrotron_vertical_profiles}
\end{figure}

Another hint is provided by Fig.~\ref{fig:B_vs_synchrotron_vertical_profiles}, were we show the
vertical profiles of $L_{1.4\mathrm{GHz}}$ and of the magnetic field. 
The vertical structure seems to be extremely similar, which tends also to disclose any other
quantities than the magnetic field to influence the
synchrotron luminosity. 
These results suggest 
that our model is not strongly dependent on the physical processes participating in the cooling of CR electrons described in Sec.~\ref{subsubsec:cooling_timescales}, that are rather sensitive to the different chemical species numerical densities.

In conclusion, there is a strong evidence that the evolution of the  
synchrotron luminosity is mostly determined by variations of the 
magnetic field. 
It is therefore probable that considering models of cosmic ray 
electrons with less, additional or other cooling processes than those 
entering the expression of $\tau_e$ (see Eq.~\ref{eq:tau}) would not 
result in huge variations in the resulting synchrotron maps. 
However, a deeper analysis of the sensitivity of the synchrotron 
emission to different CR models will be the focus of future work.

As it has been mentioned before, the choice to set 
$H = 300$~pc 
has been motivated by the fact that this value was actually the scale height at which the stellar component was initialized in the SILCC simulations (\citealt{SILCC_project_molecularclouds}). With a typical velocity for the galactic winds of $v_0 = 50$ km/s, this leads to a cooling timescale around $\tau_{\mathrm{wind}}\simeq 5$ Myr, which is 
quite short compared to typical values 
that are estimated for the Milky Way. 
Furthermore, it is also clear that the value of $\tau_{\mathrm{wind}}$ should not stay constant over time; the vertical structure will change in time which should modify both values of $H$ and $v_0$. In this sense, the vertical disc dynamics should play an important role in the simulations and they could, in principle, change the overall luminosity of the simulation box. However, the SILCC simulations are conducted over 60 Myr, which is relatively short time compared to the typical time scales of the Milky Way's dynamics, which could justify our choice to use a constant cooling time scale. 

Additionally, we have computed the synchrotron luminosity~\eqref{equ:SILCC_synch_lum_power_formula} by setting $\tau_\mathrm{wind} = 0$. The results are shown as blue curves on Fig.~\ref{fig:B_intNe_rho_L}. In absence of galactic winds, the cosmic ray number density $n_{\mathrm{CR}}$ is approximately one order of magnitude higher than our fiducial model. The synchrotron luminosity shows the same trend, but being approximately half an order of magnitude higher without wind. In this sense, we can state that any value chosen between the extreme case of 
$\tau_{\mathrm{wind}}\simeq 5$~Myr 
and $\tau_{\mathrm{wind}}=0$ would not produce significant bias in the final results of synchrotron luminosity, and that our conclusions will not suffer from this choice of parameter. Of course, a more thorough investigation must be conducted in order to study the influence of a more complex vertical disc dynamics on the resulting synchrotron emission, 
which is reserved for a future work.

\subsection{Equipartition between cosmic rays and magnetic energies}
\label{subsec:results_equipartitoin}
The assumption of energy equipartition between the magnetic
field and the cosmic ray electrons is often used in the analysis of 
synchrotron radiation. 
However, this assumption has important limitations due to the possibly 
fundamentally different temporal and spatial evolution of the two 
energy densities, see, e.g., \citet{BeckKrause2005} and 
\citet{SetaBeck2019}. 
Whereas on global galactic scales the assumption of energy 
equipartition might be reasonable, the local deviations from it on 
scales below kiloparsecs can be significant. 
Numerical simulations of the interstellar medium including 
non-equilibrium cooling and cosmic ray protons in the 
advection-diffusion approximation suggest that the CR energy density 
on scales from a few up to $500$ pc only vary by a factor of a few
\citep[see, e.g.][]{GirichidisEtAl2018a}. 
On the contrary, the magnetic field energy scales with the density 
\citep[e.g.][]{Crutcher2012, HennebelleInutsuka2019} and thus varies 
by orders of magnitudes in different regions of the simulation box. 
The additional complication of different production channels for CR 
protons and electrons in combination with the vastly different cooling
efficiencies challenges the assumption of equipartition. 

In order to examine the exact energy density distribution of the CR 
electrons and the other energy components, one needs to include accurate 
CR electron processes including the formation of primaries and secondaries, 
which is not included in the current simulations \citep[see e.g.][for a post-processing treatment of the CR electrons]{WerhahnEtAl2021a, WerhahnEtAl2021b, maria_paper_cosmic_rays} but will be covered in future simulations including spectral CRs as in \citet{GirichidisEtAl2020, GirichidisEtAl2022a}.

\begin{figure*}
    \centering
    \includegraphics[width = 0.95\textwidth]{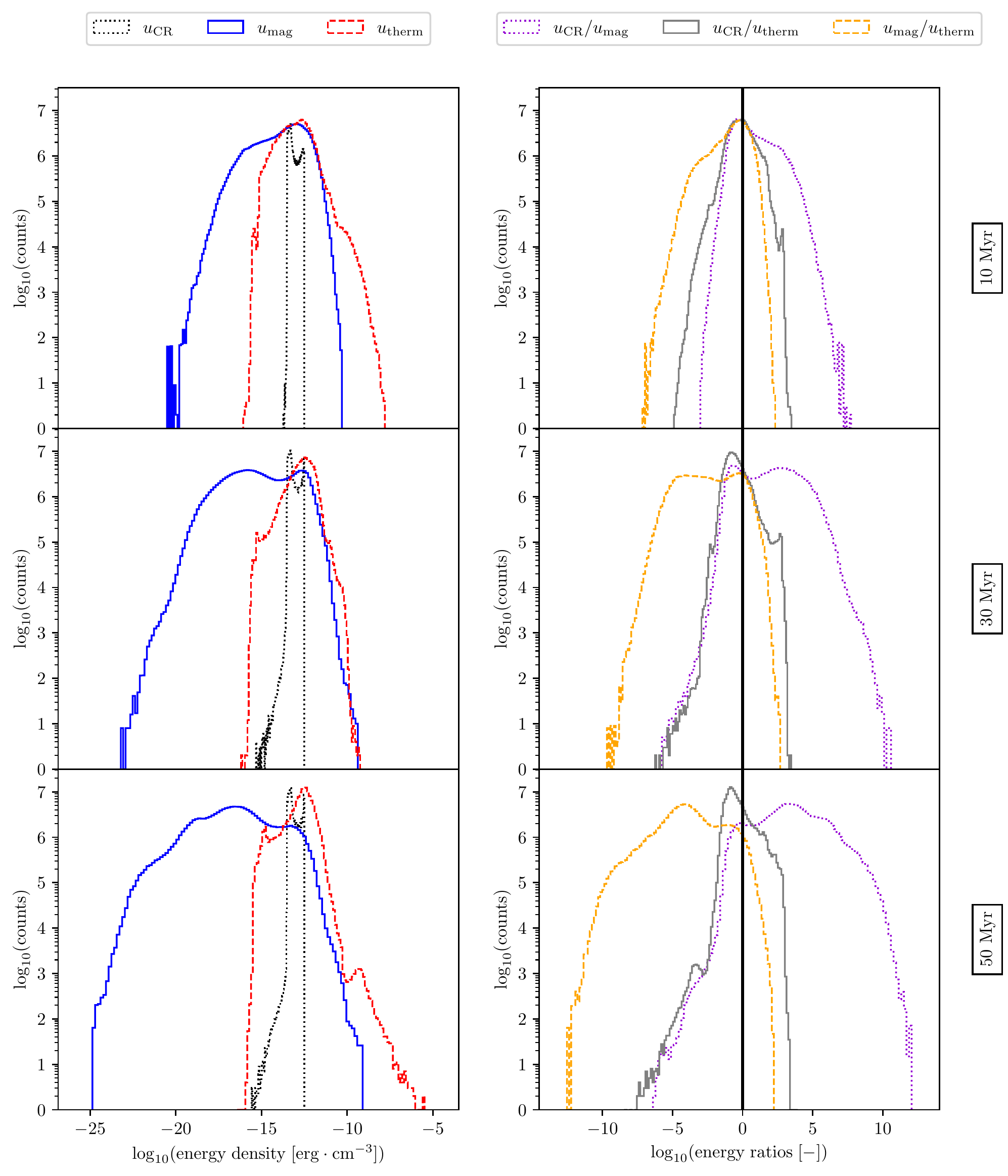}
    \caption{Distribution of CR, magnetic and thermal energy densities (left panels), and of the different ratios of those energy components (right panels)
    for the whole simulation domain, at $10$, $30$ and $50$ Myr, for $B_0 = 3$ $\mu$G.}
    \label{fig:3d_energy_density_b03}
\end{figure*}

We now want to compute the energy density of the magnetic field, the one of cosmic ray
electrons and the thermal energy explicitly from the simulations and 
test the validity of the equipartition assumption.  
The magnetic energy density is defined as
\begin{equation}\label{eq:mag_energy_dens}
u_\mathrm{mag} \equiv \frac{B^2}{8 \pi}
\end{equation}
and the cosmic ray energy density as
\begin{equation}\label{eq:CR_energy_dens}
u_\mathrm{CR} \equiv\int_{m_ec^2}^{\infty}  N_e(E)E~\mathrm{d}E,
\end{equation}
where $N_e(E)$ is given by formula \eqref{CR_e_power_spectrum}. 
The thermal energy density is given by 
\begin{equation}
u_{\mathrm{therm}} = \frac{3}{2}nk_BT,
\end{equation}
where $k_B$ is the Boltzmann constant, 
$n \equiv n_{\mathrm{neut}}+n_{\mathrm{ion}}$ the total gas density, 
and $T$ the temperature. 
In this formula, $n$ and $T$ are outputs of the simulations.
Figure~\ref{fig:3d_energy_density_b03} shows the evolution of
the three energy components, as well as the evolution of the three-dimensional distribution of the cooling timescales (given by Eqs.~\ref{ion_timescale}-\ref{wind_timescale}). In particular, the cooling timescales are calculated for two energies, 511 keV and 10 GeV.

Figure~\ref{fig:3d_energy_density_b03} 
shows the distributions of $u_{\mathrm{mag}}, u_{\mathrm{therm}}$ and $u_{\mathrm{CR}}$ (left panels), and the distributions of the ratios of those densities (right panels). 
At early times, the three energy components peak approximately at 
$10^{-13}$~erg~cm$^{-3}$ (or equivalently around
$10^{-2}$~eV~cm$^{-3}$). 
The distribution of the magnetic energy is wider than the 
other curves, which can be understood as an effect of the initial 
distribution of the magnetic field whose value decrease from the galactic 
plane following a Gaussian distribution. 
Additionally, the three different energy density ratios are peaked
around $10^{-14}$~erg~cm$^{-3}$, 
even though the curves are far from being Dirac delta functions and 
spread over almost ten orders of magnitudes.

At later times, the magnetic energy distribution is even wider, spreading over more than 15 orders of magnitude. 
This is not surprising given that, in absence of dynamo processes in the simulations, 
the magnetic energy tends to decrease 
(as it was shown of Fig.~\ref{fig:RM_magfields_evolution}).
On the other hand, the CR energy distribution also tends to get wider, 
which could be either linked to the decrease of the magnetic energy, 
to the decrease of the gas density, 
or the combination of the two effects. 
Additionally, the CR distribution has two peaks near $10^{-14}$ erg cm$^{-3}$. 

Regarding the energy density ratios, it clearly appears that the shape of the histogram of $u_{\mathrm{CR}}/u_{\mathrm{therm}}$ does not evolve over time. 
Indeed, $u_{\mathrm{CR}}$ is mainly influenced by the average gas density, that does not
change dramatically over time (see for example Fig.~\ref{fig:B_intNe_rho_L}),
and $u_{\mathrm{therm}}$ by the temperature variations that occur on small scales compared to the overall size of the box because of supernovae explosions.
Only the magnetic energy distribution undergoes a major evolution in time, which explains the shifts observed in the $u_{\mathrm{CR}}/u_{\mathrm{mag}}$ and $u_{\mathrm{mag}}/u_{\mathrm{therm}}$ curves. 
Finally, note that there seems to be no significant differences between Fig.~\ref{fig:3d_energy_density_b03}, and Fig.~\ref{fig:3d_energy_density_b06},
that displays the energy ratios for the simulation with $B_0 = 6$ $\mu$G. 

Overall, our results demonstrate that it is still difficult to 
justify or disprove the hypothesis of energy equipartition that is often adopted in 
literature.
Our work
highlights that equipartition can be reached (at least) locally, 
in the sense that in many grid cells $u_\mathrm{mag}\approx u_\mathrm{CR}$. 
However, the dynamical range in ratios and the resulting large deviations do not justify the assumption of equipartition globally.
Of course, there could be a bias introduced by our simplified modelling of the CR electrons, and by the numerical dissipation of the magnetic energy in the simulations.
None the less, we do not expect that the distribution of energy ratios will significantly narrow down to equipartition values. This suggests that the equipartition hypothesis could be possibly applied locally, but not globally in the ISM.


\section{Conclusions}
\label{sec:conclusions}

In this paper, we have analysed the magnetic field properties from magnetohydrodynamical simulations 
of the SILCC project. 
We have examined two observational quantities, 
namely the Faraday rotation measure (RM), 
which requires an assumption of the density of free electrons, 
as well as the synchrotron luminosity which depends on 
the distribution of cosmic ray (CR) electrons.

In the first part of our work, we calculated the RM along the 
three Cartesian axes, as well as its power spectrum, in order to study the 
characteristic length scales of the Faraday rotation maps. 
We tested three different models for the electron density 
(Sec.~\ref{subsubsec:FRM_ne_models}). 
In the first model a constant density is assumed 
($n_e = 10^{-3}$~cm$^{-3}$). 
The second model is based on the contribution of ionized chemical species, 
specifically the density of hydrogen (H$^{+}$) and carbon monoxide (CO$^{+}$) 
that is an output of the simulation. 
In the third model $n_e = f n$,
where $n$ is the total gas number density and $f = 10^{-3}$.

We find that the 
time evolution of the rms value of RM differs strongly
between the different electron 
density models. 
For $n_e$ model 1 (constant free electron density), the average values range from $\sim 0.1$ to $ \sim 0.5$ rad m$^{-2}$ (Fig.~\ref{fig:rm_maps}).
The models taking into account the ionisation 
degree of the chemical species show a different evolution in 
comparison to the one based on the total density of the gas. 
There, the highest RM values are 
approximately $25$ and $\sim 80$ rad m$^{-2}$. 
The $n_e$ model 3 shows even higher peaks, 
approximately around 300 rad m$^{-2}$.
We find that the high RM values at those peaks are dominated by small spatial zones of a few parsec with extremely high electron density. 
Those high-density areas in the simulations are 
created by the combination of heating of the gas in combination with compression 
shock fronts from supernovae explosions. 
Our results ultimately indicate that the strong fluctuations in the electron density 
need to be taken into account in the analysis of observational data.
The assumption of a constant electron density in calculating RM
from post-processing dynamo simulations in periodic boxes 
(see for example \citealt{faraday_rotation_SSD}), 
faces problems if applied for understanding observations of
the individual thermal phases of the ISM where the density varies over
many orders of magnitude.

For the second part of our work, we implemented the semi-analytical 
model for the power spectrum of cosmic ray electrons developed by 
\citet{schober_galactic_synchrotron_emission} and investigated projected maps of 
simulation 
outputs. 
We find that the synchrotron luminosity decreases over time following a similar temporal evolution as the magnetic intensity.
Since the CR electron density is almost constant over
time, and the average gas density does not vary
significantly, we deduce that the temporal evolution of the synchrotron luminosity is mainly determined by the evolution of the magnetic field (Figs.~\ref{fig:B_intNe_rho_L} and \ref{fig:B_vs_synchrotron_vertical_profiles}).
This is further supported by the similarity of the vertical profiles of both quantities.

Finally, we computed CR electrons, magnetic and thermal energies (namely 
$u_{\mathrm{mag}}, u_{\mathrm{CR}}$ and $u_{\mathrm{therm}}$), 
compared them, and tested the hypothesis of equipartition between the magnetic and cosmic rays energies that is vastly assumed in literature.
Our results show that, the magnetic energy density changes significantly locally as well as globally over time
(see Figs.~\ref{fig:3d_energy_density_b03},
\ref{fig:mass_energies_numbparticles_evolution} and  \ref{fig:3d_energy_density_b06}).
On the contrary, the distribution of the CR energy density varies only slightly. 
Similarly, the thermal energy density also does not show major variations in times. 
The two latter distributions peak at approximately $10^{-13}$ erg cm$^{-3}$ (corresponding approximately to $10^{-2}$ eV cm$^{-3}$). 
Regarding the ratios of the energy densities, the only curve that stays centered around unity is $u_{\mathrm{CR}}/u_{\mathrm{therm}}$, however, with large wings on both sides. 
The ratio of the magnetic to CR energy density is very broad and varies over time due to the 
dynamics in the magnetic field -- and possibly as well
due to CR effects, which are not included in the current model. 
Effectively, more than half of all regions are approximately 1 to 4 orders of magnitude away from equipartition, which does not justify equipartition to be a valid assumption.

Our work has demonstrated that extreme care is needed for the interpretation of continuum radio observations of the interstellar medium.
This concerns also the analysis of the plethora of radio data expected from the new generation of radio telescopes, above all the \textit{Square Kilometre Array} SKA\footnote{www.skatelescope.org}, especially when observing the ISM in the more distant Universe. 
Ultimately, a reliable analysis of the observables of cosmic magnetic fields are crucial for answering some of the central questions of modern astrophysics, like the nature of turbulent galactic dynamos or the propagation of cosmic rays.

\section*{Acknowledgements}
We thank Abhijit B.\ Bendre for useful comments on the manuscript.
YR and JS acknowledges the support by the Swiss National Science Foundation under Grant No.\ 185863.
PG acknowledges funding from the European Research Council under ERC-CoG grant CRAGSMAN-646955.

\section*{Data Availability}

The simulation data are publicly available at \href{http://silcc.mpa-garching.mpg.de}{http://silcc.mpa-garching.mpg.de} under data release 6 (DR6, \citealt{SILCC_project_molecularclouds}). The analysis scripts for this study will be shared upon request to the corresponding author.

\appendix

\section{Evolution of total mass, kinetic, magnetic and thermal energies}
\begin{figure}
    \centering
    \includegraphics[width = \columnwidth]{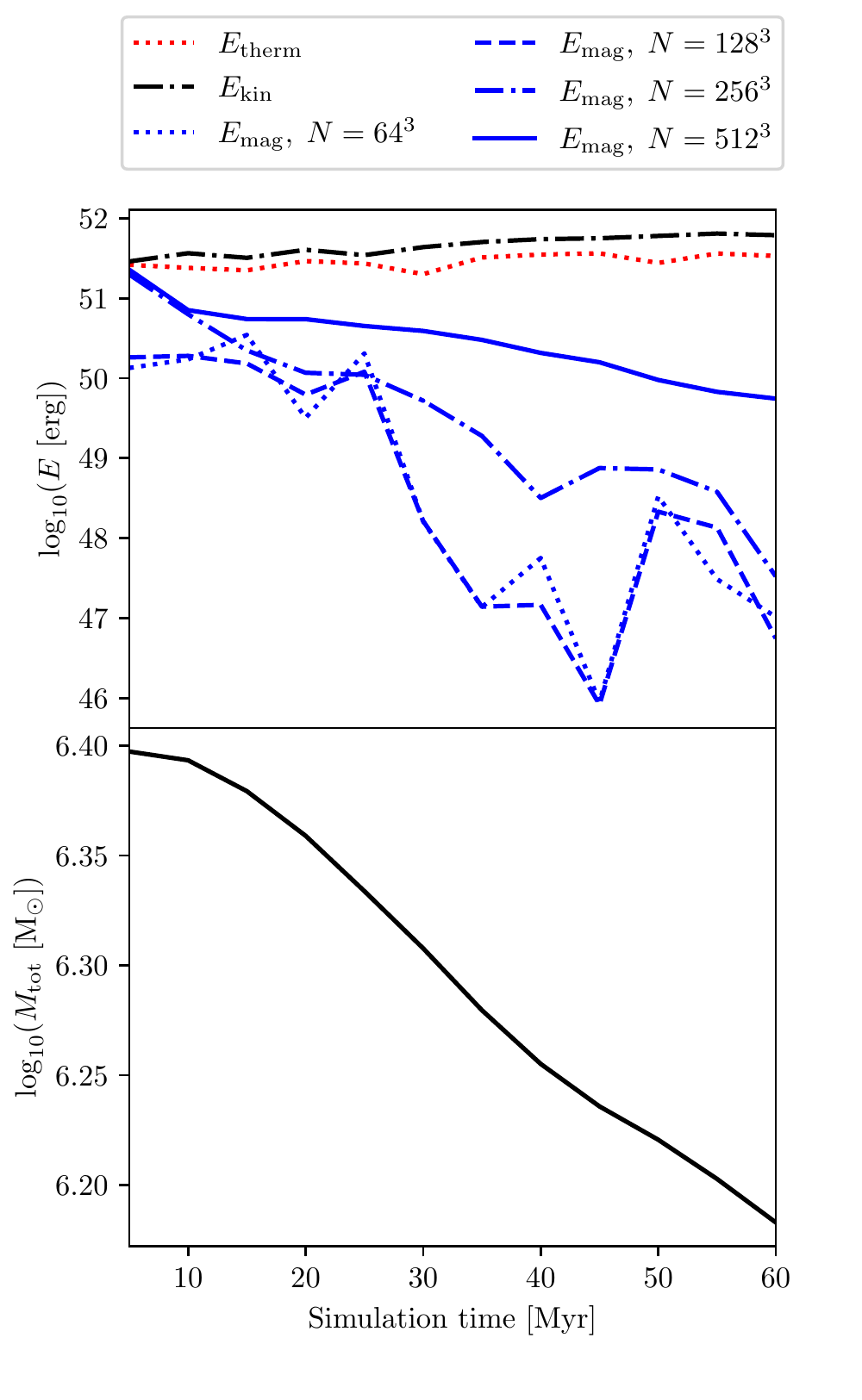}
    \caption{Evolution of overall magnetic, kinetic and thermal energies of the simulation box (top) and of the total mass of the system (bottom).
    With the exception of $E_\mathrm{mag}$, all quantities here are calculated for a resolution of $N = 512^3$.  
    $E_\mathrm{mag}$ is underestimated for low resolution, which can be seen in the top panel.
    }
    \label{fig:mass_energies_numbparticles_evolution}
\end{figure}

Figure~\ref{fig:mass_energies_numbparticles_evolution} shows 
the evolution of magnetic, thermal and kinetic energies, as well as the total 
mass of the system. 
The total mass is decreasing over time which is an due to the outflowing boundary conditions used in the simulations and the dynamics in the ISM that launches outflows from the disc.

In the top panel of Fig.~\ref{fig:mass_energies_numbparticles_evolution}
the effect of the resolution for the analysis on the total magnetic energy is shown.
It is clear that this energy is underestimated in lower resolutions ($N = 64^3, 128^3$), but this underestimation is less important in $N= 256^3$. 
Overall, the most important deviation from the maximum resolution occur between approximately 
30 and 50~Myr, 
which corresponds to the typical period of time during which small high-density molecular clouds form. 
Under-resolving the strongly contracting zones results in errors of the total magnetic energy. 

\section{Effect of varying resolution on rotation measures}
\label{appendix:varying_resolution}

\begin{figure}
    \centering
    \includegraphics[width = \columnwidth]{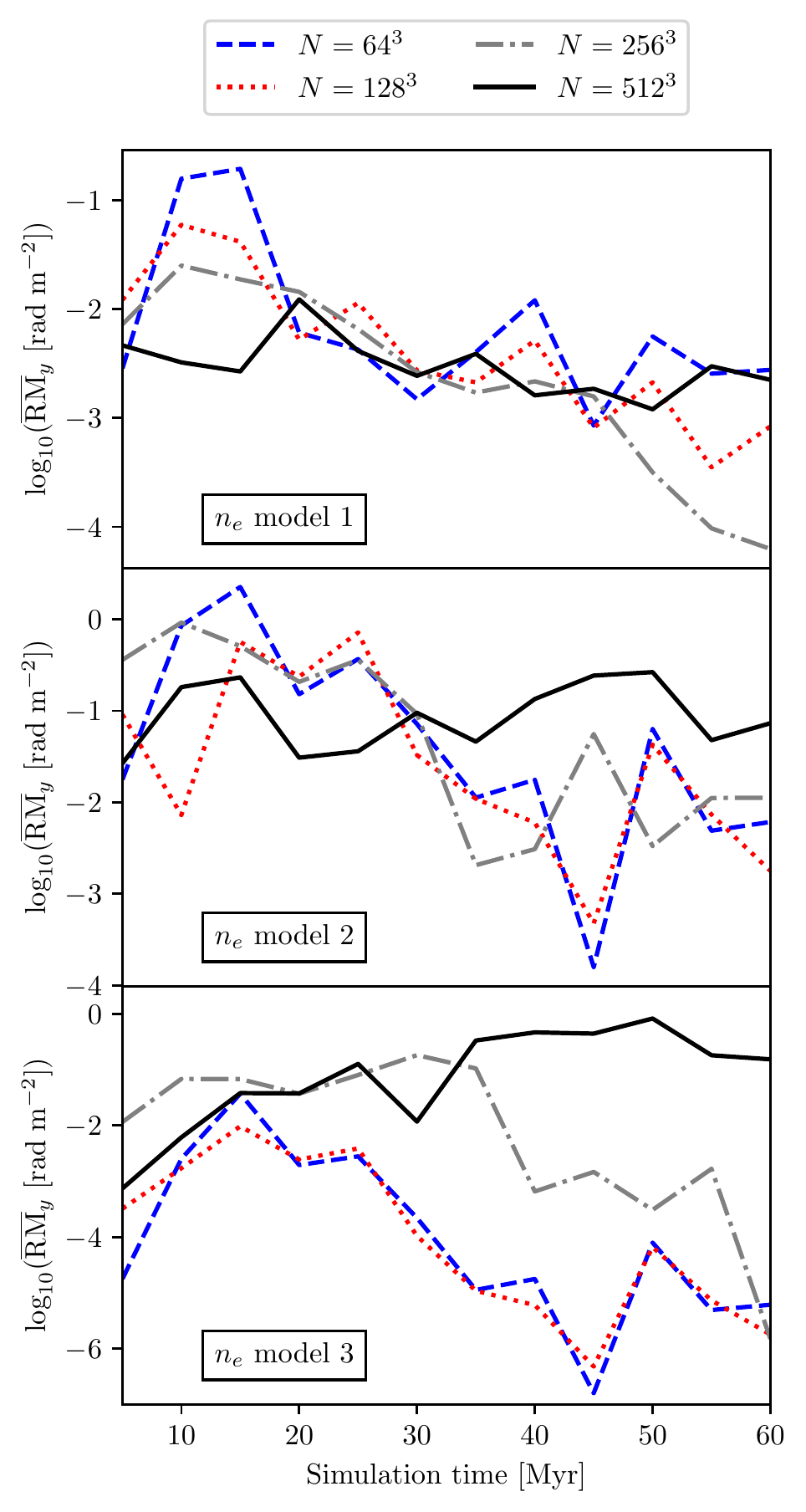}
    \caption{Evolution of the rotation measure along the $y$ axis for different resolutions of the simulations. Those curves are computed for $B_0 = 3$ $\mu$G.}
    \label{fig:varying_resolution}
\end{figure}

Figure~\ref{fig:varying_resolution} shows the effect of varying
resolution on the values of the average of the rotation measure 
calculated along the $y$ axis. 
For $n_e$ model 1 (top panel), we observe that increasing the resolution
does not result in a dramatic change of the final RM, and the 
global trend of the latter is a decreasing evolution.
Variations caused by the varying resolution are however observed and are the most
pronounced in the end of the simulation 
(approximately 1.5 orders of magnitude of difference between $N = 64^3$ and $N= 256^3$). 
Given that this $n_e$ model 1 is based on a constant free electron 
density, this indicates that the magnetic energy is not strongly 
affected by the resolution. 
As for $n_e$ model 2, we observe that the curve for $N = 512^3$ stays (almost) 
constant in time, with typical variations approximately of half an
order of magnitude, while the curves for the other resolutions are
decreasing by approximately 2 orders of magnitude. 
This model is based the density of ionized species, and then 
it is likely (although not certain) that small zones are more 
ionized than the rest of the domain because of 
supernova explosions.
Finally, the most important dependence on resolution is observed for 
$n_e$ model
3 in which the electron density is assumed to be a fraction of 
the total gas density. 
Indeed, the difference of the RM value between the curve 
for $N = 512^3$ and other resolutions is almost five orders of magnitudes at late times.
However, this is not surprising because the creation of very dense and small molecular 
regions in the simulation are not resolved in the coarse resolutions but are smoothed 
out if not considering the full AMR resolution. 
As a result, we lose a significant amount of the Faraday rotation signal. 
In general, those results tell us that we have to be extremely careful when 
considering the resolution at which the ISM chemical dynamics are modelled. 
If constant electron density as a simple model is ruled out, then a bias in the RM 
could emerge at higher resolutions. 
The results could still vary for higher resolutions in the MHD simulations as in 
\cite{Girichidis2021}.

\section{Implementation of the supernova rate in the cosmic ray model}
\label{appendix:numerical_SN_rate}
In Sec.~\ref{sec:SYNCH_emission},
the supernova rate is
used to calculate the injection of cosmic rays in Eq.~\eqref{equ_SN_rate} for each numerical grid cell.

Each cell is given a supernovae probability per unit time.
Following the implementation of the location of the supernovae explosions given in \citet{SILCC_project_molecularclouds}, we proceed as follows. Let $p_{0,\mathrm{I}}$ and $p_{0,\mathrm{II}}$ be the probabilities of having a type Ia and type II supernova in each grid cell of the central plane. We assume $p_{0,\mathrm{II}} = 5 p_{0,\mathrm{I}}$. The probability in the central plane is supposed to be uniform, so we set $p_{0,\mathrm{II}} =N_0/(N_{\mathrm{grid}})^2$, where $N_0$ is the unknown variable to determine. Enforcing that the sum of all probabilities equals unity, $N_0$ is determined by the following relation:
\begin{equation}
\begin{split}
1 = 2N_{grid}^2\left(\sum_{k = 0}^{\left\lfloor N_{\mathrm{grid}}/2\right\rfloor}p_{0,\mathrm{I}}e^{-\frac{(k\Delta z)^2}{H_{\mathrm{I}}^2}} + \sum_{k = 0}^{\left\lfloor N_{\mathrm{grid}}/2\right\rfloor}p_{0,\mathrm{II}}e^{-\frac{(k \Delta z)^2}{H_{\mathrm{II}}^2}}\right) 
\end{split}
\end{equation}
where $N_{\mathrm{grid}}$ is the resolution of the simulation, $\Delta z$ is the size of a grid cell (in pc), and $H_{\mathrm{I}}$ and $H_{\mathrm{II}}$ are the scale-height of the Gaussian distribution for type Ia and type II supernovae, respectively. 
We adopt the simulation values and use 
$H_{\mathrm{II}} = 50$~pc and $H_{\mathrm{I}} = 300$~pc. Finally, we multiply the latter equation by $15$ $\mathrm{Myr}^{-1}$ in order to obtain the supernovae rate for each grid cell.

\section{Cooling timescales in the run with $B_0 = 6$ $\mu$G}
 
\begin{figure*}
    \centering
    \includegraphics[width = 0.95\textwidth]{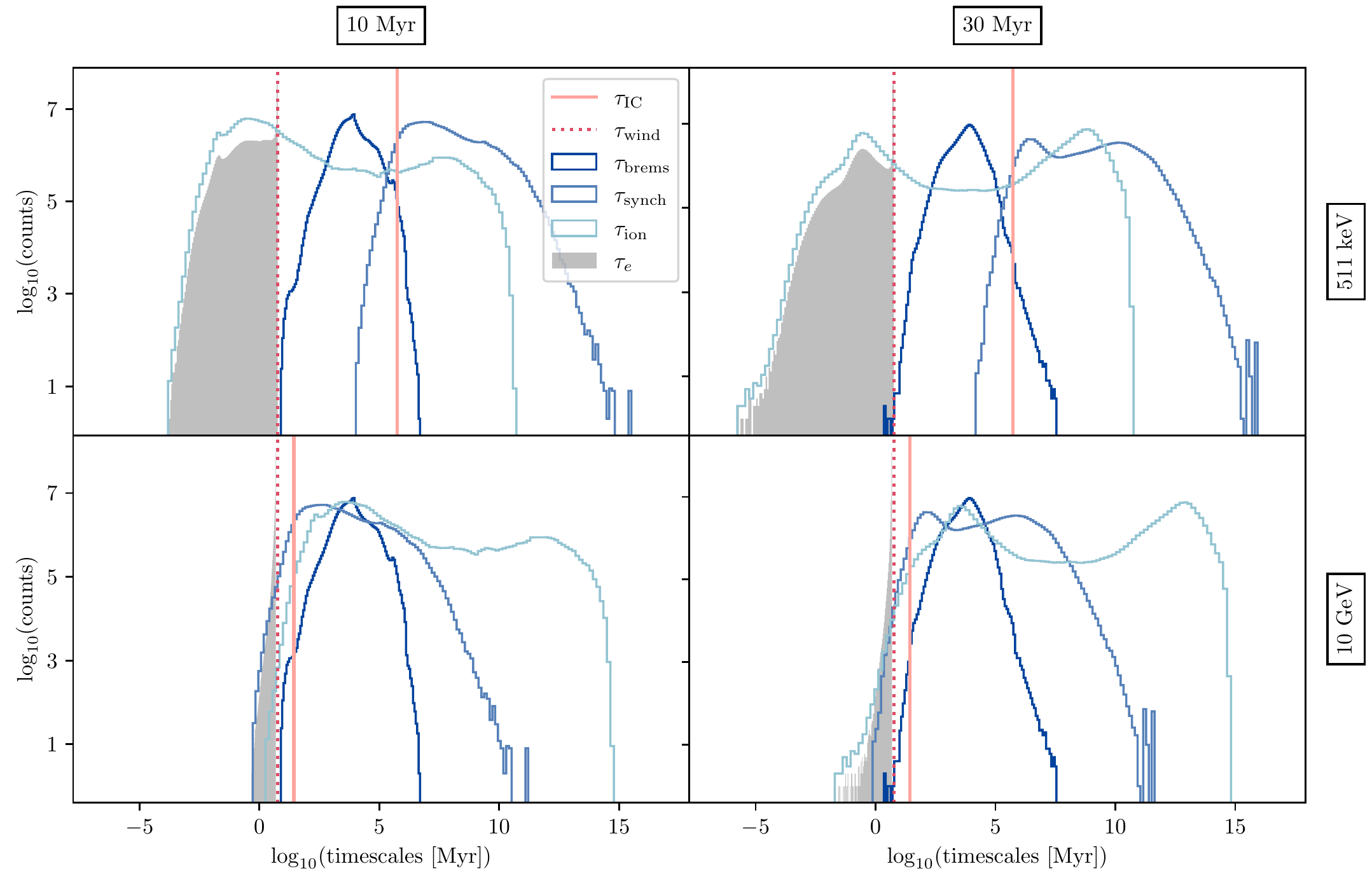}
    \caption{Histograms of the different cooling timescales involved in the expression of \eqref{eq:tau} entering the diffusion-loss equation \eqref{diffusion_loss_equation} (for the whole three-dimensional simulation box), at 10 and 30 Myr, 
    for 511 keV
    (electron mass energy) and 10 GeV, and for $B_0 = 6$ $\mu$G.}
    \label{fig:3d_cooling_timescales_b06}
\end{figure*}

Figure~\ref{fig:3d_cooling_timescales_b06} shows the cooling 
timescale, discussed in Sec.~\ref{subsubsec:cooling_timescales}, but for $B_0 = 6$ $\mu$G.

\section{Energy densities in the run with $B_0 = 6$ $\mu$G}
 
\begin{figure*}
    \centering
    \includegraphics[width = 0.95\textwidth]{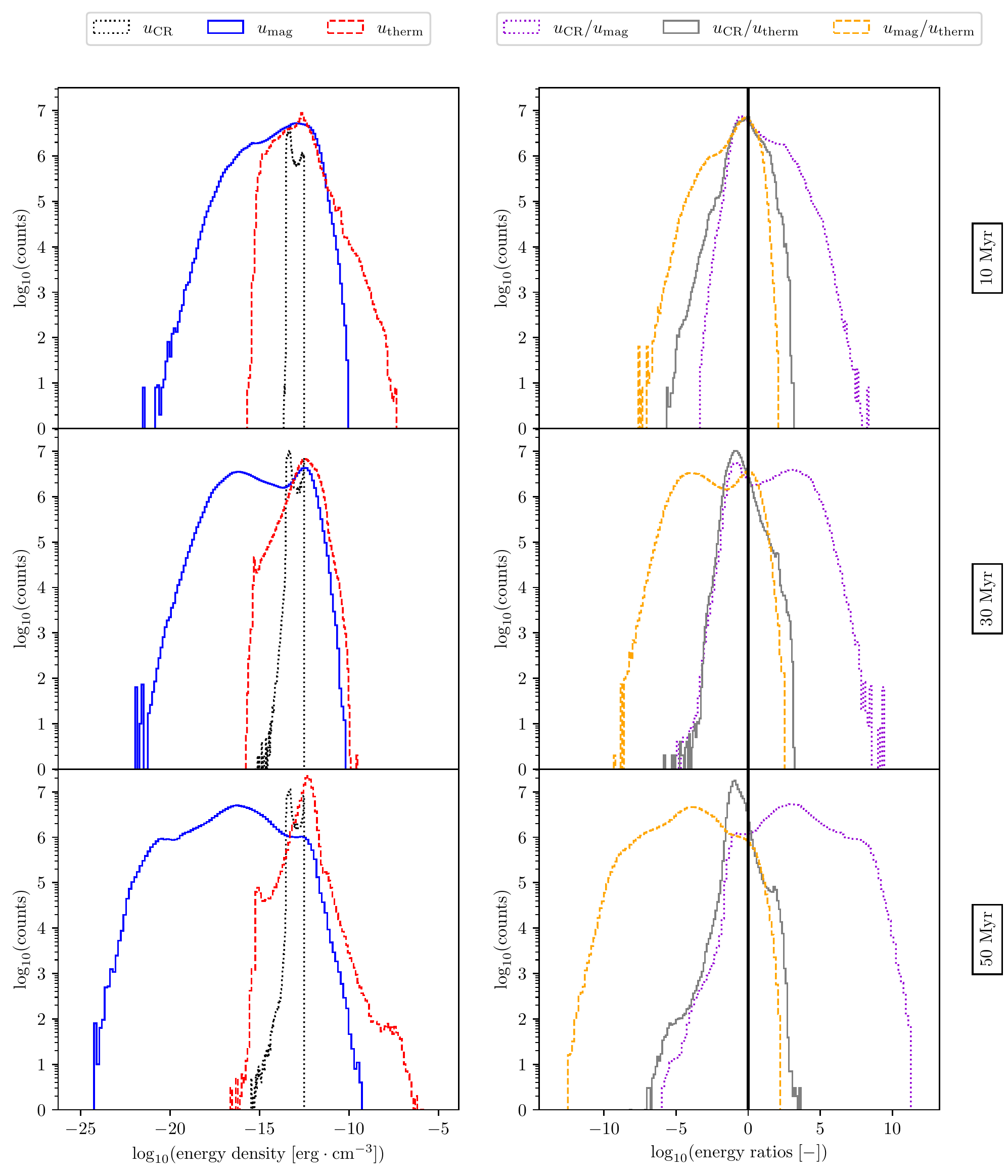}
    \caption{Distribution of CR, magnetic and thermal energy densities (left panels), and of the different ratios of those energy components (right panels) for the whole simulation domain, at 10, 30 and 50 Myr, for $B_0 = 6$ $\mu$G.}
    \label{fig:3d_energy_density_b06}
\end{figure*}

Figure~\ref{fig:3d_energy_density_b06} shows the distribution 
of thermal, magnetic and cosmic rays energy densities, discussed 
in Sec.~\ref{subsec:results_equipartitoin}, but for $B_0 = 6$ $\mu$G.

\label{lastpage}
\end{document}